\documentclass[a4paper, 11pt]{article}

\usepackage[margin=1in]{geometry}
\usepackage{setspace}
\usepackage{microtype}
\usepackage{float}

\usepackage{amsmath} \allowdisplaybreaks
\usepackage{amssymb}
\usepackage{amsthm}
\theoremstyle{definition}    %
\newtheorem{theorem}{Theorem}

\newtheorem{proposition}{Proposition}
\newtheorem{definition}{Definition}
\theoremstyle{remark}

\usepackage{natbib}
\usepackage{bibentry}

\usepackage{booktabs} %
\usepackage{graphicx}
\usepackage{appendix}
\usepackage[counterclockwise]{rotating} %
\usepackage{subcaption} %
\usepackage[normalem]{ulem} %
\usepackage{showexpl}
\usepackage{threeparttable} %

\usepackage[hyphens]{url}
\usepackage[bookmarks=false,hidelinks,breaklinks=true]{hyperref}

\usepackage{tikz}
\usetikzlibrary{automata,calc,trees,positioning,arrows,chains,shapes.geometric,%
decorations.pathreplacing,decorations.pathmorphing,shapes,%
matrix,shapes.symbols,plotmarks,decorations.markings,shadows}

\usepackage{pgf}
\pgfdeclarelayer{background}
\pgfdeclarelayer{foreground}
\pgfsetlayers{background,main,foreground}

\usepackage{authblk}

\usepackage[shortlabels]{enumitem}

\graphicspath{{figures/}}

\renewcommand{\Pr}{\mathbb{P}}  %

\newcommand{\email}[1]{{\href{mailto:#1}{\nolinkurl{#1}}}}

\usepackage{bm}

\DeclareMathSymbol{\Gamma}{\mathord}{operators}{"00}
\DeclareMathSymbol{\Delta}{\mathord}{operators}{"01}
\DeclareMathSymbol{\Theta}{\mathord}{operators}{"02}
\DeclareMathSymbol{\Lambda}{\mathord}{operators}{"03}
\DeclareMathSymbol{\Xi}{\mathord}{operators}{"04}
\DeclareMathSymbol{\Pi}{\mathord}{operators}{"05}
\DeclareMathSymbol{\Sigma}{\mathord}{operators}{"06}
\DeclareMathSymbol{\Upsilon}{\mathord}{operators}{"07}
\DeclareMathSymbol{\Phi}{\mathord}{operators}{"08}
\DeclareMathSymbol{\Psi}{\mathord}{operators}{"09}
\DeclareMathSymbol{\Omega}{\mathord}{operators}{"0A}

\DeclareMathOperator*{\argmax}{arg\,max}

\usepackage{xcolor}

\usepackage{etoolbox}
\makeatletter
\def\do#1{\@namedef{#1c}{\ensuremath{\mathcal{#1}}}}
\docsvlist{A,B,C,D,E,F,G,H,I,J,K,L,M,N,O,P,Q,R,S,T,U,V,W,X,Y,Z}
\makeatother

\def\Rb{\mathbb{R}}

\newcommand{\dd}{\mathop{}\!\mathrm{d}}
\newcommand{\dx}{\mathop{}\!\mathrm{d}x}

\newcommand{\dt}{\mathop{}\!\mathrm{d}t}
\newcommand{\dr}{\mathop{}\!\mathrm{d}r}
\newcommand{\dw}{\mathop{}\!\mathrm{d}w}
\newcommand{\dv}{\mathop{}\!\mathrm{d}v}

\renewcommand{\bar}[1]{\mkern 1.5mu\overline{\mkern-1.5mu#1\mkern-1.5mu}\mkern 1.5mu}

\usepackage{framed}

\usepackage{fancyvrb,listings}
\title{Dynamic Pricing of an Expiring Item under Strategic Buyers with Stochastic Arrival}

\author[1]{Suyeon Choi}
\author[1]{Changhyun Kwon}
\author[2]{Seungki Min\footnote{Corresponding Author: \email{mskyt@snu.ac.kr}}}
\affil[1]{Department of Industrial and Systems Engineering, KAIST, Daejeon, 34141, Republic of Korea}
\affil[2]{Business School, Seoul National University, Seoul, 08826, Republic of Korea}

\begin{document}
\maketitle

\begin{abstract}
\noindent
\textbf{Problem definition:}
We study the optimal dynamic pricing for an expiring ticket or voucher, sold by a time-sensitive seller to strategic buyers who arrive stochastically with random private values.
The item's expiring nature creates a fundamental conflict: the seller’s pressure to sell before expiration leads to price reductions, which in turn incentivizes buyers to wait for a better deal.
We seek to find the seller's optimal pricing policy that resolves this tension.

\noindent
\textbf{Methodology/results:}
The primary analytical challenge is that a buyer's type is two-dimensional (defined by both private valuation and arrival time), which makes the equilibrium intractable under general strategies.
To address this, we introduce the \emph{Value-Based Threshold (VBT) strategy}, a tractable framework that decouples these two dimensions.
Using this framework, we establish the necessary and sufficient conditions for equilibrium via an \emph{ordinary differential equation} and provide a formal proof of its existence and a numerical procedure of its construction.
We then apply this equilibrium characterization to derive near-optimal pricing policies for two stylized market regimes: a constant price for thin markets and a linear discount for thick markets.
Our numerical frontier analysis validates these theoretical benchmarks and provides refined insights into the optimal pricing structure.

\noindent
\textbf{Managerial implications:} 
Our findings offer a clear resolution to the conflict between the need for a quick sale and buyers' strategic waiting.
A seller under pressure to sell early, either due to operating in a thick market with intense competition or having high personal time sensitivity, is best served by a \emph{linear discounting schedule}.
In a thin market with little competition, a \emph{constant price} is optimal to neutralize a single buyer's powerful incentive to wait for a better deal; our analysis also shows this simple policy is remarkably robust across a wide range of market conditions. 
For a patient, time-insensitive seller, a \emph{quasi-auction schedule} that maintains a high price until a final, sharp drop is most effective for maximizing revenue by aggregating demand.

\end{abstract}

\section{Introduction}
In recent years, the \emph{resale of items with an expiration date}, most notably event tickets, has grown rapidly across online marketplaces \citep{courty2019ticket}.
Unlike a conventional retail environment where firms sell multiple products in bulk, these markets are characterized by individual sellers offering a single indivisible item with an expiration date through online platforms.
Markets for expiring items present the distinctive characteristic that the item becomes worthless upon expiration, making it unsellable.
Furthermore, the expiration of the item incentivizes prospective buyers to delay their purchases in anticipation of potential price reductions.
Such strategic buyer behavior is particularly pronounced in secondary markets where items are traded via online auctions throughout the selling horizon such as eBay \citep{backus2015sniping}.
Within such market structures, pricing strategy plays a crucial role in maximizing the seller's utility \citep{waisman2021selling}.
This context motivates a central research question: \emph{what is the optimal pricing policy for a monopolist selling a single, expiring item to strategic buyers who arrive stochastically}?

The seller in this market undergoes a distinct \emph{temporal pressure} stemming directly from the nature of the expiring item.
Unlike goods that may retain residual value, the terminal value of an expiring item is zero, creating the risk of a total economic loss for an unsold unit.
This characteristic makes the seller highly time-sensitive and provides a strong incentive for a rapid sale.
We formalize this core feature of the market through a time-sensitive utility function.

Conversely, the seller's temporal pressure, by inducing price declines as the expiration approaches, creates an incentive for prospective buyers to engage in \emph{strategic delaying} behavior \citep{cachon2009purchasing, aflaki2020becoming}.
This behavior is empirically supported by observations that secondary market prices for event tickets tend to decrease as the event approaches, fostering a consumer segment of ``bargain hunters'' who wait for last-minute deals \citep{drayer2009value}.
In stark contrast to the seller, we posit that buyers are not time-sensitive.
The utility from items such as tickets is derived at the time of consumption, not purchase, rendering the timing of the acquisition itself irrelevant to a buyer's utility beyond price considerations.
This modeling choice aligns with prior work on ticket pricing \citep{sweeting2012dynamic, jiaqi2019designing, cachon2024enigma}.
The primary trade-off for a buyer, therefore, is between securing a lower price by waiting and the increasing risk of stockout due to competition from other buyers, whose arrivals we model as a stochastic process.

To analyze the strategic interaction between a time-sensitive seller and patient, strategic buyers, we formulate the problem as a Stackelberg game.
This framework is appropriate as the seller (the leader) first commits to a pricing schedule over a finite horizon.
Buyers (the followers), upon their stochastic arrival, observe this schedule and subsequently make their optimal purchasing decisions.
Our analysis makes two primary contributions within this framework.
First, we formally characterize the buyers' equilibrium behavior.
Second, we derive near-optimal, closed-form pricing schedules for the seller under stylized market conditions, providing both theoretical insights and actionable managerial guidance.

\paragraph{A tractable framework for buyer equilibrium.}
Our first contribution is to develop a tractable framework for analyzing the buyer equilibrium.
In our model, a buyer's strategic decision is driven by private information which is inherently two-dimensional: their \emph{private valuation} and their \emph{stochastic arrival time}.
The former governs their willingness to pay, while the latter dictates their window of opportunity to purchase.
The interdependence of these two dimensions is a well-known difficulty in dynamic mechanism design \citep{bergemann2006dynamic, pavan2014dynamic, bergemann2011survey}, leading to complex strategy spaces where equilibrium behavior is often intractable.
To overcome this challenge, we introduce and analyze a class of strategies, termed \emph{Value-Based Threshold strategies}.
The core of this approach is a threshold time function that maps a buyer's valuation to a target purchase time, thereby decoupling the two components of the buyer's type.

This structure allows us to establish a formal procedure for proving the existence and uniqueness of an equilibrium among buyers in response to a committed price schedule.
We first derive a set of \emph{necessary conditions}, including a key ordinary differential equation, that any equilibrium threshold function must satisfy. These conditions serve as a constructive tool, allowing us to derive a unique candidate equilibrium strategy for any given pricing schedule.
Subsequently, we establish a set of \emph{sufficient conditions} that act as a verification tool.
These conditions confirm whether the derived candidate strategy is indeed a stable equilibrium by ensuring no buyer has a unilateral incentive to deviate.

Notably, \citet{briceno2017optimal} also introduce effectively the same notion of strategies. However, their formal equilibrium existence result is limited to the discrete valuation setting, where buyers take on one of two types—high or low. For a setting with a continuous distribution of buyer valuations, they assumed the existence of an equilibrium without a formal proof.
In contrast, our work directly addresses this limitation. Together, the necessary and sufficient conditions we establish provide a systematic method to formally prove the existence of an equilibrium in a continuous valuation setting. This provides a rigorous foundation for analyzing such strategies in more general and realistic market environments.

\paragraph{Optimal pricing schedules for stylized markets.}

Our second contribution leverages our equilibrium framework to derive actionable pricing schedules for the seller. To obtain tractable and insightful results, we analyze two stylized market regimes defined by the buyer arrival rate: thin markets with infrequent arrivals and thick markets with frequent arrivals.

Our analysis first addresses the \emph{thin market} regime, where a low arrival rate results in negligible competition among buyers.
In such a sparse market, an arriving buyer holds a significant strategic advantage.
With virtually no risk of being preempted, they have a dominant incentive to wait for the lowest possible price if the schedule is not constant.
The seller's primary challenge is thus to design a policy that neutralizes this incentive to delay.
We show that the optimal solution is a \emph{constant pricing schedule} in this preemptive setting.
By removing the prospect of any future discount, this flat price forces an immediate ``take-it-or-leave-it'' decision upon the buyer's arrival.
This constant price policy, which we also prove induces a unique equilibrium for strategic buyers, is therefore the seller's most effective tool to compel an early transaction.

At the other extreme, our analysis addresses the \emph{thick market} regime, where a high arrival rate fosters intense competition among buyers.
In this environment, the significant risk of being preempted by a rival outweighs the potential benefit of strategic delay. This compels buyers to behave as if they are myopic, purchasing immediately once the price meets their valuation.
This simplification of buyer behavior reduces the seller's problem to a direct trade-off between price and time, for which the optimal policy is a \emph{linear discounting schedule}.
Intuitively, this schedule acts as a continuous auction, scanning downwards through buyer valuations at a constant rate.
This rate is determined solely by the seller's own time sensitivity, as the uniform valuation distribution provides a constant incentive for the seller to wait at any price level.
Crucially, we then establish the robustness of this policy by proving that it also induces a unique equilibrium in the full model with fully strategic buyers, confirming its implementability.

Finally, our analytical results are validated through numerical simulations, which confirm the near-optimality of these canonical policies in their respective extreme regimes. Furthermore, our analysis goes beyond these stylized cases to provide a more complete picture of the optimal strategy. The simulations reveal a clear transition driven by the seller's own time sensitivity: the optimal schedule systematically shifts from an end-loaded (quasi-auction) for patient sellers, to a flat (constant) price for moderately sensitive sellers, and finally to a front-loaded (linear) discount for impatient sellers.

\subsection{Related Works}

Over many years, optimal pricing for strategic buyers has been extensively studied.
\citet{stokey1979intertemporal} pioneered the theoretical foundations of dynamic pricing with strategic buyers, establishing a framework for intertemporal price discrimination by a monopolist over a finite horizon with unlimited inventory.
Extending these insights, \citet{besanko1990optimal} conducted a game-theoretic analysis based on subgame perfect equilibrium, demonstrating that price skimming remains effective against strategic buyers and deriving the optimal pricing trajectory over a finite horizon.
These pioneering studies bridged dynamic pricing with game-theoretic analysis by formalizing it as a dynamic game, such as Stackelberg game, thereby enabling equilibrium characterization of optimal strategies \citep{gallego2014dynamic}.

Building on these foundational works, the literature has increasingly focused on optimal dynamic pricing under \emph{strategic buyer behavior} over the decades.
Considering strategic buyers, several studies have examined how to find the optimal pricing policy under various settings, such as differences in capacity of inventory, time of pricing, demand arrival process and buyer's valuation distribution \citep{anderson2003wait,chen2009dynamic, liu2013dynamic, liu2015optimal, cachon2015price}.

One of the central interests lies in designing optimal mechanisms \citep{horner2011managing, pavan2014dynamic, board2016revenue, wu2022optimal}.
Alongside these theoretical advances, empirical studies have increasingly used real-world data to validate theoretical predictions and analyze strategic buyer behavior in actual markets \citep{waisman2021selling, gisches2021strategic}.

Due to the high cost and limited capabilities of \emph{real-time} price adjustments, research on strategic buyers throughout the 2010s focused on revenue maximization under static or discrete pricing policies.
However, with the advancement of digital infrastructure and the proliferation of online marketplaces, real-time price adjustment has become feasible, even for individual sellers.
This shift has led to a growing body of research on \emph{continuous} pricing strategies for single-item settings, reflecting the evolving dynamics of modern online marketplaces \citep{spann2024algorithmic}.

Parallel research has extended models by incorporating \emph{uncertainty} in buyer behavior.
Some studies introduce valuation uncertainty to reflect demand heterogeneity \citep{levina2009dynamic,su2009model,swinney2011selling}, while others model uncertainty in arrival process to capture the stochastic nature of online marketplaces \citep[e.g.,][]{feldman2016online,gershkov2018revenue,mashiah2023learning}.
For an extensive overview, refer to \citet{gonsch2013dynamic}.

In the same line, we study a continuous pricing schedule considering strategic buyers with stochastic arrival, which results in two-dimensional buyer types: private valuation and arrival time.
Its nature leads to complex strategy spaces, where equilibrium behavior generally depends on both dimensions and lacks a tractable representation.
Prior studies have recognized this challenge in various domains such as dynamic pricing and mechanism design \citep{bergemann2006dynamic, pavan2014dynamic, bergemann2011survey}.

\citet{briceno2017optimal} address this complexity by assuming that buyers follow a non-increasing threshold strategy, thereby reducing the dimensionality of the strategic problem.
However, their formal equilibrium existence result is limited to the \emph{discrete} valuation setting, where buyers take on one of two types—high or low.
While they adopt an optimal control framework to derive the seller’s revenue-maximizing pricing schedule, they \emph{assumed} the existence of an equilibrium in the \emph{continuous} valuation setting.

In contrast, we formally establish the existence of an equilibrium in the continuous valuation setting.
Our framework thus offers a practical procedure for equilibrium verification in environments with continuously distributed buyer types.
Moreover, we introduce a unique utility framework that captures the real-world dynamics of expiring items.
This departs fundamentally from the standard dynamic pricing literature, which typically incorporates buyers’ time-dependent disutility, as seen in \citet{briceno2017optimal}.

Our model assigns time-dependent disutility exclusively to the seller, while buyers remain temporally neutral.
The prevailing literature typically incorporates time-dependent disutility for buyers, assuming their patience diminishes as time progresses.
Even in studies addressing perishable goods, both the seller and buyers are often symmetrically subject to time discounting \citep{besanko1990optimal, chen2009dynamic, osadchiy2010selling, liu2013dynamic}.

However, such an assumption fails to capture the distinctive dynamics of expiring items in online marketplaces, where time pressure is largely borne by the seller. 
As shown by \citet{drayer2009value} and further discussed in \citet{drayer2012dynamic, cachon2024enigma}, buyers in secondary ticket markets do not necessarily experience disutility over time, even though price adjustments are made dynamically as the event nears. 
In fact, \citet{drayer2012dynamic} emphasized that dynamic ticket pricing contexts primarily reflects seller-side time pressure rather than buyer impatience.
By explicitly assigning time-dependent disutility only to the seller, our framework better captures the economic reality of expiring items in online markets and offers a structurally distinct alternative to conventional dynamic pricing models.

\subsection{Paper Organization}
The remainder of this paper is organized as follows. Section~\ref{sec:problem} formally sets up our market model, defining the problems faced by both the seller and the buyers. In Section~\ref{sec:equilibrium}, we introduce our tractable equilibrium framework, the Value-Based Threshold Strategy, and establish the necessary and sufficient conditions that prove its existence. Section~\ref{sec:optimal} applies this framework to derive the near-optimal pricing schedules for thick and thin markets and presents our numerical analysis. Finally, Section~\ref{sec:conclusion} concludes the paper.
All the proofs are provided in Appendix~\ref{app:proof}.

\section{Problem Setup} \label{sec:problem}
In this section, we formally introduce our model, which frames the sale of a single expiring item as a Stackelberg game.
The seller, acting as the leader, commits to a public pricing schedule at the outset of the market.
Subsequently, multiple strategic buyers arrive stochastically and, as followers, compete to acquire the item by choosing their optimal purchase times.
The following subsections provide a detailed characterization of the market environment and the respective optimization problems faced by the seller and the buyers.

\subsection{The Market Model} 
We consider a market consisting of a single seller and multiple buyers, in which a single indivisible, expiring item is sold over a finite time horizon denoted by $\Tc = [0, T]$.
The item's value becomes zero upon expiration at time $T$.
The strategic interaction unfolds chronologically as follows.

At the outset of the sales horizon, $t=0$, the seller acts as the leader by committing to a public pricing schedule, $p: \Tc \to \Rb_+$.
This schedule, assumed to be \emph{continuous} and \emph{non-increasing}, dictates the price $p(t)$ for the entire sales horizon and is observed by all potential buyers upon their arrival.

Throughout the sales horizon, $t \in [0,T]$, buyers arrive according to a Poisson process with rate $\lambda$.
Each arriving buyer $i$ is characterized by a privately observed type $(v_i, \alpha_i)$, where $v_i$ is their valuation drawn independently from a uniform distribution over $\Vc=[0,1]$, and $\alpha_i$ is their arrival time.

Upon their arrival at time $\alpha_i$, each buyer acts as a follower. They observe the complete pricing schedule $p(\cdot)$ and, based on their private information, strategically determine an optimal \emph{target purchase time}, $\tau_i$.
The item is sold and the market concludes at the \emph{sale time} $\tau^* = \min_i \tau_i$, which is the earliest of all buyers' target purchase times.
The transaction occurs at a price of $p(\tau^*)$, as dictated by the seller's committed schedule.
If no purchases are made by time $T$, the item expires unsold.

It is common knowledge that buyers' valuations are drawn from the uniform distribution and that they arrive according to a Poisson process with rate $\lambda$.
The seller's pricing schedule $p(\cdot)$ also becomes public information, fully observable to any buyer upon their arrival.
In contrast, each buyer's realized type, their specific valuation $v_i$ and arrival time $\alpha_i$, is private information, unknown to the seller and other buyers.
This asymmetric information structure implies that buyers make decisions under uncertainty about the presence and types of their competitors.

\subsection{Buyer's Decision Problem}
Upon arrival at time $\alpha_i$, a buyer's private valuation $v_i$ is realized. From this moment forward, the buyer faces a continuous-time optimal stopping problem.
At any given time $t \geq \alpha_i$, they must decide whether to purchase the item at the current price $p(t)$ or to continue waiting.
Waiting offers the potential for a lower future price but carries the risk that a competing buyer will purchase the item first.

While this continuous decision-making process accurately describes the buyer's situation, a more tractable and equivalent formulation exists.
Because the pricing schedule is deterministic and no new market information is revealed after the buyer's arrival, the problem can be simplified without loss of generality.
The setup is equivalent to one where the buyer's full type, $(v_i, \alpha_i)$, is known to them at $t=0$, and they pre-commit to a single \emph{target purchase time},
\begin{equation}
    \tau_i \in [\alpha_i, T] \cup \{ \infty \}.
\end{equation}
This target time represents their intended purchase moment, contingent on the item's availability, with $\tau_i = \infty$ signifying no intent to purchase (e.g., if the price $p(t)$ remains above the buyer's valuation $v_i$ for all $t \in \Tc$).
This perspective allows us to model the buyer's complex dynamic problem as a single choice of $\tau_i$.

The item is sold to the buyer with the earliest target purchase time, 
\begin{equation} \label{eq:sale_time}
    \tau^* = \min_i \{ \tau_i \},
\end{equation}
at a price of $p(\tau^*)$.
If no buyers arrive or if all buyers choose $\tau_i = \infty$, $\tau^*$ is consequently defined as infinity.
We assume that simultaneous purchases do not occur, ensuring that $\tau_i \ne \tau_j$ for any $i \ne j$; this is a mild assumption that is endogenously satisfied by the equilibrium strategies we analyze later.

Each buyer chooses their target time $\tau_i$ to maximize their \emph{interim utility}, defined as the potential payoff multiplied by the probability of a successful purchase (calculated at time $t=0$):
\begin{equation}\label{eq:buyer_utility}
    \big( v_i - p(\tau_i) \big) \Pr( \tau_i = \tau^*),
\end{equation}
Here, $\Pr( \tau_i = \tau^*) $ is the probability that the buyer $i$'s intended purchase time is the earliest among all present and future competing buyers.
Based on their expectations regarding the presence and strategies of other buyers, a buyer of a given type chooses the target purchase time that maximizes this interim expected utility.

\subsection{Seller's Decision Problem}

Our model captures the seller's urgency, which arises from the item's expiration.
At the end of the sales horizon, the item loses all value, making the seller highly sensitive to the timing of the sale.
To reflect this, we define the seller's objective as maximizing an expected utility function composed of two components: the expected revenue from the sale and a \emph{linear} disutility from the time spent waiting for the sale to occur.

The seller chooses a pricing schedule $p(\cdot)$ to maximize her expected utility, $U^s$, defined as:
\begin{equation} \label{eq:seller_utility}
    U^s = \bar{r}_s - \beta \bar{\tau}_s,
\end{equation}
where $\bar{r}_s$ is the expected revenue, $\bar{\tau}_s$ is the expected waiting time, and $\beta \in \mathbb{R}_+$ is the seller's time-sensitivity coefficient, representing the marginal disutility from waiting.
The expected revenue is the price at the time of sale, conditional on a sale occurring before expiration: 
\begin{equation}
    \bar{r}_s = \mathbb{E} \Big[ p(\tau^*) \mathbb{I}\{ \tau^* < \infty \} \Big],
\end{equation}
where $\mathbb{I}\{ \cdot \}$ is the indicator function.
The expected waiting time is the sale time if the item is sold, and the full horizon length $T$ if it expires unsold:
\begin{equation}
    \bar{\tau}_s= \mathbb{E}\Big[ \min\{\tau^*, T\} \Big].
\end{equation}

The seller's problem is to find the optimal pricing schedule $p^*(\cdot)$ that maximizes this expected utility by resolving a fundamental trade-off.
A more aggressive discounting schedule may reduce the expected waiting time $\tau_s$ but will likely decrease expected revenue $r_s$, and vice versa.
Within the Stackelberg framework, this constitutes the leader's problem. The seller must determine her schedule in anticipation of the rational responses of the buyers, as their equilibrium behavior is itself a function of the chosen price path.

\section{Equilibrium Analysis} \label{sec:equilibrium}
This section characterizes the equilibrium that arises among buyers for any given pricing schedule committed by the seller. 
The competition for a single item, coupled with each buyer's incomplete knowledge of their rivals' types, defines a dynamic game of incomplete information. 
Accordingly, we formalize the interaction among buyers as a Bayesian game.

\subsection{General Buyer Strategy and Equilibrium}
Within our game-theoretic framework, each buyer, after observing their type $(v_i, \alpha_i)$, chooses a target purchase time $\tau_i$.
This choice is governed by a strategy, $\sigma$, which is formally defined as a mapping from the buyer's type space to their action space, $\sigma: \Vc \times \Tc \to \Tc \cup \{ \infty \}$.
The strategy $\sigma(v, \alpha)$ prescribes the target purchase time $\tau$ for any given buyer type, subject to the feasibility constraint that a buyer cannot purchase before arriving, i.e., $\sigma(v, \alpha) \geq \alpha$.

The buyer's objective is to select a target purchase time that maximizes their interim utility.
Given all other buyers follow a specific strategy profile $\sigma$, the utility of a buyer of type $(v, \alpha)$ choosing a time $\tau \in [\alpha, T] \cup \{\infty\}$ is given as
\begin{equation} \label{eq:general_interim}
    \Pi^b(\tau; v, \alpha, \sigma,p) 
    = \mathbb{E}
    \bigg[ 
       (v - p(\tau)) \ \mathbb{I} 
        \Big\{ 
	    	\tau < \min_{i \in \{1, 2, \ldots, N_-\}} \sigma(v_i, \alpha_i) 
		\Big\} 
    \bigg],
\end{equation}
which is obtained by formalizing the success probability $\Pr(\tau_i = \tau^*)$ in \eqref{eq:buyer_utility}.
Here, the probability is taken over the random realizations of the competing buyers. 
Specifically, this includes the number of other buyers, $N_- \sim \text{Poisson}(\lambda T)$; their private valuations, $v_i \sim \text{Uniform}(\Vc)$; and their arrival times, $\alpha_i \sim \text{Uniform}(\Tc)$.

An equilibrium is a state where no player has a unilateral incentive to deviate.
In our analysis, we focus on a \emph{symmetric Nash equilibrium}, which is a strategy $\sigma^*$ that is a best response to itself. Formally, it must satisfy:
\begin{equation}\label{eq:equilibrium}
	\sigma^*(v,\alpha)
	= \inf 
        \argmax_{\tau \in [\alpha,T] \cup \{\infty\} } \Pi^b(\tau;v,\alpha,\sigma^*,p),
        \quad \forall v \in \Vc, \quad \forall \alpha \in \Tc.
\end{equation}
The use of the infimum ensures the earliest target time is selected in cases where multiple times might yield the same maximal utility.

\subsection{Value-Based Threshold (VBT) Strategy} \label{subsec:VBT_def}

The general strategy $\sigma(v,\alpha)$ defined in the previous section poses significant analytical challenges.
The joint dependence on both valuation and arrival time leads to a complex strategy space that makes the characterization of an equilibrium intractable in most cases.
To address this challenge, we restrict our attention to a structured class of strategies that is both analytically tractable and intuitively appealing.

\begin{definition}[Value-based threshold strategy] \label{defn:vbt}
    A buyer's strategy $\sigma$ is a \emph{value-based threshold (VBT) strategy} induced by a \emph{threshold time function} $w: \Vc \rightarrow \Tc$, if it satisfies:
    \begin{equation} \label{eq:vbt}
        \sigma(v, \alpha) = \max \{ w(v), \alpha \},
        \quad \forall v \in \Vc,  \quad \forall \alpha \in \Tc.
    \end{equation}
\end{definition}

The core idea behind the VBT strategy is to \emph{decouple} the two dimensions of the buyer's type.
A buyer first determines an ``ideal'' purchase time, $w(v)$, based solely on their valuation.
If they arrive at the market before this ideal time (i.e., $\alpha < w(v)$), they wait until their target moment $w(v)$ to purchase.
If they arrive after their ideal time has already passed (i.e., $\alpha > w(v)$), they purchase immediately upon arrival.
The threshold time function $w(v)$ thus represents a buyer's strategic patience, which is dictated by their willingness to pay.

To ensure this threshold time function is analytically well-behaved, we introduce a set of regularity conditions.
We denote the set of all such regular threshold time functions by $\Wc$.

\begin{definition}[Regularity of threshold time functions]
A threshold time function $w : \Vc \rightarrow \Tc$ is said to be \emph{regular} (i.e., $w \in \Wc$) if, 
    \begin{enumerate}[label=(\roman*)]
        \item \label{cond:w_monotonicity} \( w \) is non-increasing on $\Vc$;
        \item \label{cond:w_strict_monotonicity} \( w \) is strictly decreasing on \( [\underline{v}_w, \overline{v}_w] \cap \Vc \);
        \item \label{cond:w_continuity} \( w \) is continuous on \( (\underline{v}_w, \overline{v}_w] \cap \Vc  \); and 
        \item \label{cond:w_differentiability} \( w \) is right-differentiable on \( [\underline{v}_w, \overline{v}_w)  \cap \Vc \),
    \end{enumerate}
    where $\underline{v}_w := \inf\{ v : w(v) < \infty \}$ and $\overline{v}_w := \sup\{ v : w(v) > 0 \}$.
\end{definition}

\begin{figure}
    \centering
    \includegraphics[width=1.0\linewidth]{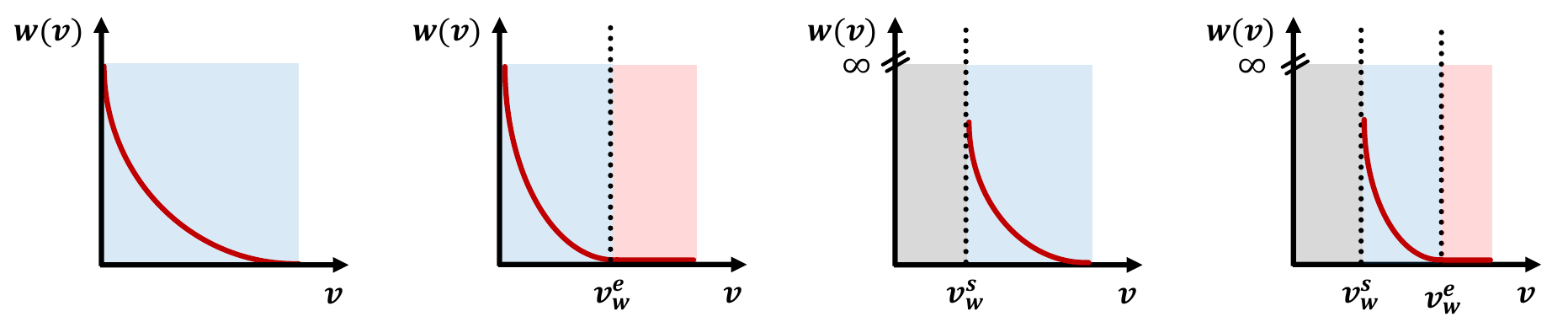}
    \caption{Representative forms of the threshold time function $w(v)$. 
    Buyers with valuations below the active range ($v < \underline{v}_w$), shown in the gray-shaded region, never intend to purchase ($w(v)=\infty$). Conversely, those with valuations above this range ($v>\overline{v}_w$), shown in the red-shaded region, intend to purchase immediately upon arrival ($w(v)=0$).}
    \label{fig:w_examples}
\end{figure}

The valuation cutoffs $\underline{v}_w$ and $\overline{v}_w$ partition the buyer population into up to three distinct behavioral groups.
Buyers with valuations below this range ($v < \underline{v}_w$) never intend to purchase; this typically corresponds to scenarios where the terminal price $p(T)$ remains strictly positive, thereby excluding buyers whose valuations fall below this price.
Those with valuations above this range ($v > \overline{v}_w$) intend to purchase immediately upon arrival.
The central group consists of buyers with ``active'' valuations ($v \in [\underline{v}_w, \overline{v}_w]$), who strategically wait for their specific target time.
As illustrated in Figure~\ref{fig:w_examples}, the different panels depict how the presence and combination of these groups can vary depending on the specific form of the threshold time function $w$.

These regularity conditions serve distinct purposes. The non-increasing condition~\ref{cond:w_monotonicity} is interpreted in terms of strategic patience; it requires that a buyer with a higher valuation be no more patient than one with a lower valuation, which is intuitive as they have more to lose from a stockout.
Within the active range, the strictly decreasing condition~\ref{cond:w_strict_monotonicity} is a stronger requirement that prevents ties by ensuring each valuation maps to a unique target time.
Finally, the conditions of continuity~\ref{cond:w_continuity} and differentiability~\ref{cond:w_differentiability} are technical requirements that ensure the function is sufficiently well-behaved for the subsequent equilibrium analysis to be tractable.

By restricting our analysis to this class of regular VBT strategies, we achieve a crucial simplification.
The original, intractable problem of finding a general two-dimensional strategy function, $\sigma(v,\alpha)$, is now reduced to a more manageable problem: finding a one-dimensional, regular threshold time function, $w(v)$.
It is this reduction in dimensionality that makes the equilibrium analysis tractable, as we will demonstrate in the following section.

\subsection{Characterization of VBT Equilibrium} \label{subsec:VBT_equilibrium}

Having defined the Value-Based Threshold (VBT) strategy, we now proceed to formally characterize the resulting equilibrium. This analysis unfolds in three steps.
First, we derive the explicit payoff functions for both buyers and the seller.
Second, we establish the necessary conditions that any equilibrium must satisfy, which provides a method for constructing a candidate equilibrium.
Finally, we provide the sufficient conditions needed to verify that such a candidate is indeed a stable symmetric Nash equilibrium.

\subsubsection{Payoffs under a Symmetric VBT Strategy}

In this section, we derive the explicit payoff functions for both buyers and the seller under the assumption that all buyers adopt a symmetric VBT strategy.
The foundational step is to determine the probability that the item remains unsold over time, which is the key to all subsequent payoff calculations.

\begin{proposition}[Survival probability] \label{prop:VBT_survival}
    Suppose that all buyers are playing the VBT strategy $\sigma_w$ induced by a regular threshold time function $w \in \Wc$.
    The probability that the item remains unsold at time $t$ is given by:
    \begin{equation}
        \Pr( \tau^* > t ) = e^{-\lambda t\big(1-w^{-1}(t)\big)}, \quad \forall t \in \Tc,
    \end{equation}
    where
    \begin{equation} \label{eq:w_inverse}
    w^{-1}(t) := \inf \{ v \in \Vc : w(v) \leq t \}.
    \end{equation}
\end{proposition}

This result stems from the fact that a sale occurs before time $t$ only if a buyer with a high valuation $v \geq w^{-1}(t)$ arrives.
Without the VBT framework, calculating this survival probability would be analytically intractable, requiring a complex calculation involving the inverse of the general two-dimensional strategy function $\sigma(v, \alpha)$.
The VBT strategy's key contribution is to decouple value and time, which simplifies this complex problem and yields the closed-form solution in the proposition.
This closed-form survival probability is, in turn, the key to explicitly defining the buyer's interim utility.

\begin{proposition}[Buyer's interim utility] \label{prop:VBT_buyer}
   Under the same assumption as in Proposition \ref{prop:VBT_survival}, the interim utility of a buyer of type $(v,\alpha)$ choosing a target purchase time $\tau \in [\alpha,T]$ is given by
    \begin{equation} \label{eq:VBT_interim}
        \Pi^b( \tau; v, \alpha, \sigma_w, p) 
        = \Pi^b( \tau; v, 0, \sigma_w, p) 
        = (v-p(\tau)) e^{-\lambda \tau \big(1-w^{-1}(\tau)\big)}, \quad \forall \tau \in [\alpha,T].
    \end{equation}
\end{proposition}

The significance of Proposition \ref{prop:VBT_buyer} is that it transforms the intractable expectation from the general model \eqref{eq:general_interim} into a tractable, closed-form expression for the buyer's interim utility.
This simple, differentiable function is independent of the buyer's arrival time $\alpha$, which greatly simplifies the subsequent best-response analysis.
This approach also allows for an explicit characterization of the seller's payoffs, which we present next.

\begin{proposition}[Seller's expected utility]\label{prop:VBT_seller}
    Suppose that all buyers are playing the VBT strategy $\sigma_w$ induced by a regular threshold time function $w \in \Wc$.
    If the pricing schedule $p$ is differentiable and non-increasing, the seller's expected revenue, $\bar{r}_s$, is given by
    \begin{equation}
        \bar{r}_s = p(0) - p(T) \cdot e^{-\lambda T (1 - p(T))} + \int_0^T e^{-\lambda t(1 - w^{-1}(t))} \cdot p'(t) \, \dt,
    \end{equation}
    and, the seller's expected waiting time, $\bar{\tau}_s$, is given by
    \begin{equation}
        \bar{\tau}_s = \int_0^T e^{-\lambda \cdot t (1-w^{-1}(t))} \dt.
    \end{equation}
\end{proposition}

These propositions provide the explicit payoff functions for all agents.
The seller's total expected utility, $U^s = \bar{r}_s - \beta \bar{\tau}_s$, is now fully characterized by the integral forms in Proposition \ref{prop:VBT_seller}.
The ability to express these payoffs in a tractable form is a direct consequence of the VBT framework, making an otherwise complex problem amenable to analysis.
With these functions formally established, we are now equipped to derive the conditions under which a VBT strategy constitutes a symmetric Nash equilibrium in the following section.

\subsubsection{Necessary Conditions and Equilibrium Construction}

We now characterize the threshold time function $w$ that induces a VBT equilibrium.
The following theorem formally states the necessary conditions that any such equilibrium function must satisfy.
These conditions provide the fundamental structure, including a key differential equation, that  governs the equilibrium.

\begin{theorem}[Necessary conditions for a VBT equilibrium] \label{thm:VBT_necc}
Suppose that the pricing schedule $p$ is differentiable and non-increasing.
If a regular threshold time function $w \in \Wc$ induces a VBT equilibrium with respect to pricing schedule $p$, then it must satisfy the following conditions: 
 \begin{enumerate}[label=(\roman*)]
        \item \label{cond:necc_no_purchase}
            For any $v \in [0,p(T)), w(v) = \infty$.
        \item \label{cond:necc_boundary}
            For $v = p(T)$, $w(v) = \inf\{t: p(t) = p(T)\}$.
        \item \label{cond:necc_ode}
            For any $v \in [p(T),1)$, the right derivative $w'(v)$ satisfies:
            \begin{equation} \tag{$*$} \label{eq:w_ode}
                w'(v) = \frac{\lambda  w(v) (v - p(w(v)))}{ \lambda (1-v) (v-p(w(v))) + p'(w(v)) }.
            \end{equation}
        \item \label{cond:necc_rationality}
            For any $v \in [p(T),1]$, $p(w(v)) \leq v$. 
\end{enumerate}
\end{theorem}

The necessary conditions of Theorem \ref{thm:VBT_necc} are presented for distinct buyer groups, which are partitioned by the terminal price, $p(T)$.
As the lowest price offered during the sales horizon, the terminal price serves as a natural boundary, separating buyers based on their fundamental willingness to participate in the market.

Conditions \ref{cond:necc_no_purchase} and \ref{cond:necc_rationality} together establish basic rationality constraints.
For buyers with valuations below the terminal price where $v<p(T)$, Condition \ref{cond:necc_no_purchase} states that a purchase is never profitable, so their target time is infinite.
For any participating buyer, Condition \ref{cond:necc_rationality} ensures the purchase price does not exceed their valuation.

Condition \ref{cond:necc_boundary} defines the behavior of the marginal buyer at the boundary for which $v=p(T)$ holds.
For this buyer, the maximum possible utility is zero, achievable at any time $t$ when $p(t)=p(T)$.
Given this indifference, our equilibrium definition in \eqref{eq:equilibrium} requires the buyer to select the earliest of these optimal times, as formalized by the condition.

Condition \ref{cond:necc_ode} captures the core dynamic of the equilibrium for the buyers with valuations above the terminal price where $v \geq p(T)$.
The ordinary differential equation \eqref{eq:w_ode}, is derived from the first-order condition of the buyer's interim utility maximization problem, using the result of Proposition~\ref{prop:VBT_buyer}.
The ODE is expressed in terms of the right derivative, the existence of which is guaranteed by Condition \ref{cond:w_differentiability} of the regularity definition.
Intuitively, it represents a delicate indifference condition: at their target time $w(v)$, an active buyer must be indifferent between purchasing immediately and waiting an infinitesimal moment longer for a slightly lower price, optimally balancing this gain against the increased risk of being preempted by a rival.

These necessary conditions also provide a direct economic interpretation for the valuation cutoffs, $\underline{v}_w$ and $\overline{v}_w$, introduced in Section \ref{subsec:VBT_def}.
The terminal price, $p(T)$, effectively becomes the lower cutoff for market participation, meaning that in equilibrium, $\underline{v}_w =p(T)$.
The upper cutoff, $\overline{v}_w$, is the valuation at which the target waiting time becomes zero for which $w(v)=0$ holds, and its value is endogenously determined by the solution to the ordinary differential equation \eqref{eq:w_ode}.
The theorem thus formally defines the boundaries of the three distinct buyer groups.

\paragraph{Equilibrium construction procedure.}
The necessary conditions from Theorem \ref{thm:VBT_necc} are not merely descriptive; they provide a constructive procedure for deriving a candidate equilibrium threshold function, $w$, for any given pricing schedule, $p$.
The procedure is as follows:
\begin{enumerate} %
    \item \textbf{Set the no-purchase region:} Set $w(v) \gets \infty$ for all $v \in [0,p(T))$.
    \item \textbf{Set the boundary condition:} Set $w(p(T)) \gets \inf\{t: p(t)=p(T)\}$.
    \item \textbf{Solve the ODE:} Starting from the initial point $v=p(T)$, solve the ordinary differential equation \eqref{eq:w_ode} for $w(v)$ over the active range of valuations until either $w(v)$ reaches zero or $v$ reaches one.
    \item \textbf{Set the immediate-purchase region}: In step 3, if the solution reaches zero at a valuation $\overline{v}_w < 1$, then set $w(v) \gets 0$ for all $v \in [\overline{v}_w, 1]$.
\end{enumerate}

This procedure provides a unique candidate threshold function for any given price schedule.
We employ this construction method in Section~\ref{subsec:frontier} to compute the threshold functions for our numerical analysis.
It remains to be verified, however, whether a candidate function derived from this procedure constitutes an equilibrium. The following section provides the sufficient conditions for this verification.

\subsubsection{Sufficient Conditions and Equilibrium Verification}

The necessary conditions from Theorem \ref{thm:VBT_necc} yield a candidate equilibrium function but do not guarantee it constitutes a stable equilibrium.
We therefore introduce the following sufficient conditions to verify that the candidate function represents a utility-maximizing choice for every buyer, ensuring no profitable unilateral deviations exist.

\begin{theorem}[Sufficient conditions for a VBT equilibrium] \label{thm:VBT_suff}
    Suppose that the pricing schedule $p$ is differentiable and non-increasing.
    A regular threshold time function $w \in \Wc$ that satisfies the following two conditions induces a VBT equilibrium:
    \begin{enumerate}[label=(\roman*)]
        \item It satisfies all the necessary conditions specified in Theorem \ref{thm:VBT_necc}.
        \item For each $v \in [p(T),1)$, the set $\argmax_{\tau \in [0,T]} (v-p(\tau)) e^{-\lambda \tau \bigl(1-w^{-1}(\tau)\bigr)}$ is connected.
    \end{enumerate}
\end{theorem}

The first condition is a prerequisite, ensuring the candidate function has the fundamental structure required of any equilibrium.
The second condition is the crucial verification step.
It confirms that the target time $w(v)$ derived from the first-order condition in Theorem \ref{thm:VBT_necc} indeed corresponds to a unique, utility-maximizing choice for every active buyer, thus ensuring the stability of the equilibrium. 

However, this second condition is not always guaranteed to hold. 
In particular, a pricing schedule with multiple distinct price drops can cause the buyer's interim utility function to develop multiple local maxima.
In such cases, the VBT framework may break down.
A buyer's optimal strategy could become dependent on their arrival time in addition to their valuation; for example, an early-arriving buyer might find it optimal to purchase at the first price drop, while a later-arriving buyer with the same valuation would prefer to wait for the second.
Characterizing such a hybrid equilibrium presents a significant analytical challenge and is a promising direction for future research.

In our numerical analysis in Section \ref{subsec:frontier}, we employ this theorem to verify that the nontrivial pricing schedules we examine do in fact induce a valid VBT equilibrium.
Together, Theorems \ref{thm:VBT_necc} and \ref{thm:VBT_suff} provide a complete toolkit for both constructing and verifying the equilibrium.

\section{Optimal Price Schedule for Two Stylized Markets} \label{sec:optimal}

Building on the characterization of the buyer's equilibrium, this section solves the seller's problem of finding the optimal pricing schedule.
To derive tractable and insightful results, we analyze two stylized market regimes that represent opposite extremes of market density: a \emph{thin market} with rare arrivals and a \emph{thick market} with frequent buyer arrivals.
For each regime, our analysis proceeds in two steps.
First, we introduce a simplifying assumption that makes the seller's optimization problem solvable, yielding a candidate optimal pricing schedule.
Second, we prove that this schedule is robustly implementable by showing it induces a unique VBT equilibrium for fully strategic buyers.
Finally, we validate our theoretical findings through a numerically efficient frontier analysis.

\subsection{The Thin Market Regime} \label{subsec:thin}
We first analyze the thin market regime, characterized by a high buyer arrival rate ($\lambda \to 0$).
In this setting, the probability of more than one buyer arriving during the sales horizon is negligible.
Without competition, an arriving buyer faces virtually no risk of being preempted and thus has a powerful incentive to delay their purchase to secure the lowest possible price.
The seller's primary challenge, therefore, is to design a policy that forces an immediate ``take-it-or-leave-it'' decision.

To formally analyze this extreme case, we introduce a \emph{preemptive market} assumption.
This simplification posits that the first-arriving buyer holds an exclusive, uncontested right to purchase the item at any point during the sales horizon.
This effectively removes all competitive pressure and isolates the strategic interaction between the time-sensitive seller and a single, patient buyer.
Under this preemptive assumption, the seller's optimization problem again becomes tractable.

\begin{theorem}[Optimal pricing in a preemptive market]\label{thm:thin_market}
    Suppose that the market is preemptive so that the first‐arriving buyer can occupy the market without immediate purchase.  
    Let $c^\star(\lambda,\beta,T)$ be the unique solution to the implicit equation:
    \[
        c^\star = 1-\frac{1}{\lambda T}\ln\left(1+\lambda T\bigl(c^\star +\tfrac{1}{2}\beta T\bigr)\right).
    \]
    Among all differentiable and non-increasing price schedules, the \emph{constant schedule}
    \[
        p(t) = \max\{ 0, \min\{ c^\star, 1 \} \}
    \]
    maximizes the seller's expected utility. Moreover, the unconstrained optimal price $c^\star$ is strictly increasing in the arrival rate $\lambda$ and the horizon $T$, and strictly decreasing in the time sensitivity $\beta$.
\end{theorem}

Theorem \ref{thm:thin_market} establishes that the optimal policy in a market dominated by a single, unopposed buyer is a constant price.
The intuition is clear: any price decline would be fully exploited by the buyer, who would simply wait for the minimum.
A constant price removes the incentive to delay and forces an immediate ``take-it-or-leave-it'' decision, which is critical for the time-sensitive seller. 
However, this result relies on the strong preemptive assumption. The following proposition addresses this by proving that a constant price schedule also induces a unique, stable VBT equilibrium in the main strategic model.

\begin{proposition} \label{prop:constant_eq}
    Any constant pricing schedule of the form $p(t) = c$ for $c \in \Vc$ induces a unique symmetric VBT equilibrium.
    This equilibrium is induced by the threshold time function:
    \begin{equation}
        w(v) = \left\{ \begin{array}{ll}
            \infty, & v < c, \\
            0, & v \geq c.
        \end{array} \right.
    \end{equation}
    Furthermore, in this equilibrium, the seller's expected revenue $\bar{r}_s$ and expected waiting time $\bar{\tau}_s$ are given by:
    \begin{equation} \label{eq:constant_eq_seller_utility}
        \bar{r}_s = c(1-e^{-\lambda T (1-c)})
        , \quad
        \bar{\tau}_s = \frac{1}{\lambda (1-c)} (1-e^{-\lambda T (1-c)}).
    \end{equation}
\end{proposition}

Together, Theorem \ref{thm:thin_market} and Proposition \ref{prop:constant_eq} provide a robust justification for the constant price schedule in thin markets.
While Theorem \ref{thm:thin_market} derives the optimal constant price $c^\star$ under the preemptive assumption, it is a powerful result that this same price level also maximizes the seller's utility in the full strategic model, using \eqref{eq:constant_eq_seller_utility}, among all possible constant price levels.
The proposition then guarantees that this optimal constant price is implementable, as it induces a unique and intuitive equilibrium: buyers with a valuation greater than or equal to the price purchase immediately, while others do not.

Furthermore, the comparative statics from Theorem \ref{thm:thin_market} provide sensible economic insights, showing that the optimal price $c^\star$ increases with buyer arrival rate and sales horizon, but decreases as the seller becomes more time-sensitive.
As the preemptive assumption becomes an increasingly accurate approximation in very thin markets, there results suggest that the constant pricing schedule is not just a viable policy, but a robust, near-optimal one.

We clarify that this paper does not claim that the equilibrium must take the VBT form for any arbitrary pricing schedule.
Rather, our contribution is to show that for the two simple and practical pricing policies we analyze --- linear and constant schedules --- a unique symmetric VBT equilibrium is guaranteed to exist.
This result establishes that these canonical policies are implementable in a market with fully strategic buyers.
Therefore, our analysis validates them as robust, near-optimal heuristics for sellers operating in thick and thin market extremes.

\subsection{The Thick Market Regime} \label{subsec:thick}
We now turn to the thick market regime, characterized by a low buyer arrival rate ($\lambda \to \infty$).
In such a market, the intense competition renders any strategic delay unprofitable.
For any individual buyer, the risk of being preempted by a rival with a higher valuation is so significant that waiting for a lower price becomes a futile strategy.
This extreme competitive pressure compels buyers to behave as if they are \emph{completely impatient}.

We formally define complete impatience as a behavior where a buyer, upon arrival at time $\alpha$, purchases the item immediately if their valuation meets or exceeds the current price, $v \geq p(\alpha)$.
This behavior is perfectly captured within our VBT framework by a specific threshold function: the inverse of the price schedule, $w=p^{-1}$. This is because an impatient buyer's latest acceptable purchase time, $w(v)$, is precisely the moment the price falls to their valuation level, $p(t)=v$.
This simplification makes the seller's optimization problem analytically tractable, allowing us to derive the optimal schedule.

\begin{theorem}[Optimal pricing under complete impatience]\label{thm:thick_market}
    Suppose all buyers exhibit complete impatience, meaning their strategy is induced by the threshold function $w = p^{-1}$ (i.e., $p(w(v))=v$ for $v \in [p(T),1]$).
    Among all differentiable and non-increasing price schedules, the \emph{linear schedule}
    \[
      p(t)=\max\{\,1-\beta\,t,\,0\}
    \]
    maximizes the seller's expected utility $U^s$.
\end{theorem}

Theorem \ref{thm:thick_market} provides a powerful and elegant result: the optimal policy in a market with completely impatient buyers is a simple linear discount, with the rate of discount determined solely by the seller's own time sensitivity.
However, this result relies on the strong assumption of myopic buyer behavior.
It does not, by itself, guarantee that this linear schedule is an implementable policy for fully strategic buyers.
The following proposition addresses this gap by proving that a linear schedule does indeed induce a unique, stable VBT equilibrium

\begin{proposition} \label{prop:linear_eq}
    Any linear pricing schedule of the form $p(t) = \max\{ b - m t, 0 \}$ for $b \in \Vc$ and $m \in \mathbb{R}_+$ induces a unique symmetric VBT equilibrium.
\end{proposition}

Proposition \ref{prop:linear_eq} is crucial as it confirms that the linear schedule is an implementable policy that induces a stable equilibrium for fully strategic buyers.
Together, Theorem \ref{thm:thick_market} and Proposition \ref{prop:linear_eq} provide a robust justification for the linear schedule in thick markets.
Theorem \ref{thm:thick_market} establishes its optimality in an idealized extreme (complete impatience), while Proposition \ref{prop:linear_eq} guarantees its implementability in the main strategic model.
This two-step argument provides a justification for the linear schedule as a near-optimal policy.

The structure of this near-optimal schedule is highly intuitive.
The optimal slope of the price decline is a direct reflection of the seller's time sensitivity, $\beta$.
This direct link arises because the thick market assumption simplifies the seller's strategic problem.
In a general setting, a seller must design a price schedule to both extract revenue and manage buyers' strategic waiting.
However, in a thick market, intense competition accomplishes this incentive management for the seller.
With strategic delay rendered ineffective, the seller's complex task is reduced to a more fundamental trade-off between their own revenue and waiting time, making their optimal policy a direct reflection of their personal time sensitivity.
The schedule's linearity, in turn, is a direct consequence of the uniform valuation assumption; with a non-uniform distribution, the schedule would become non-linear to adapt to the changing density of valuations.

As the complete impatience assumption becomes an increasingly accurate approximation for markets with high arrival rates, we conclude that the simple linear schedule \( p(t) = \max\{ 1 - \beta t, 0 \} \) serves as a robust, near-optimal policy for thick markets.

\subsection{Efficient Frontier Analysis} \label{subsec:frontier}

We now validate our theoretical findings and explore the performance of different pricing policies through a numerical analysis.

\begin{figure}
    \centering
    \begin{subfigure}[b]{0.32\linewidth}
        \includegraphics[width=\linewidth]{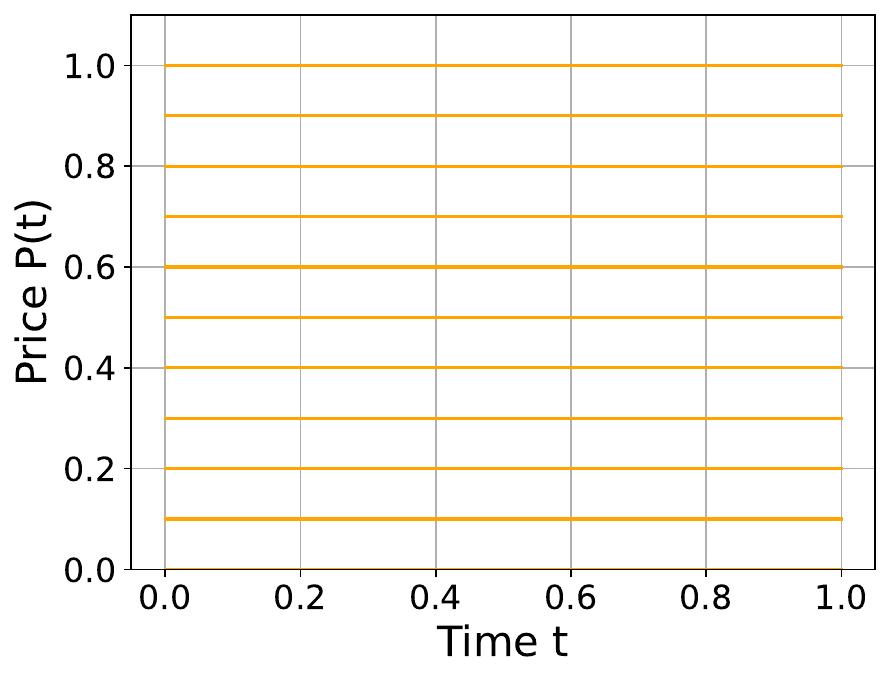}
        \captionsetup{justification=centering}
        \caption{
            Constant\\[0.5em]
            $p(t) = c$
        }
        \label{fig:sub3}
    \end{subfigure}
    \hfill
    \begin{subfigure}[b]{0.32\linewidth}
        \includegraphics[width=\linewidth]{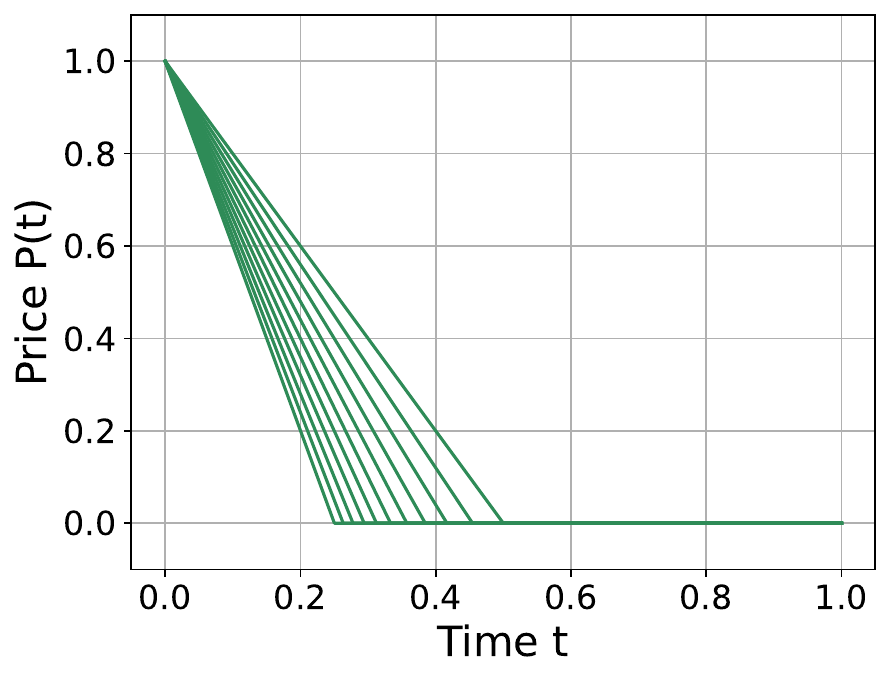}
        \captionsetup{justification=centering}
        \caption{
            Linear\\[0.5em]
            $p(t) = \max\{1 - mt, 0\}$
        }
        \label{fig:sub2}
    \end{subfigure}
    \hfill
    \begin{subfigure}[b]{0.32\linewidth}
        \includegraphics[width=\linewidth]{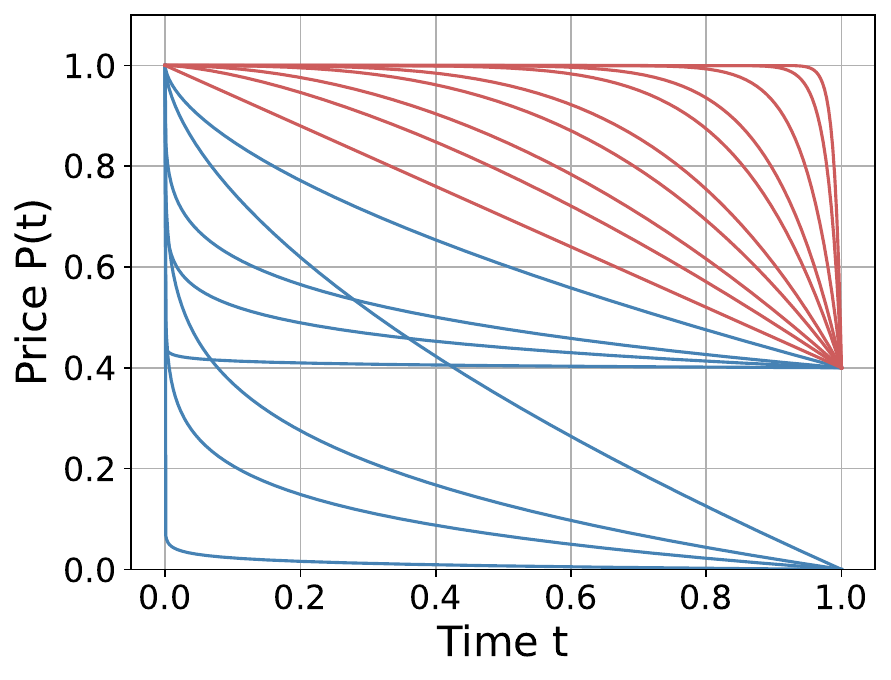}
        \captionsetup{justification=centering}
        \caption{
            Polynomial/quasi-auction\\[0.5em]
            $p(t) = 1 - (1 - r)t^\alpha$
        }
        \label{fig:sub1}
    \end{subfigure}
    \caption{The parameterized pricing schedule families evaluated in the numerical analysis.}
    \label{fig:pricing_schedules}
\end{figure}

\paragraph{Setup.}
In our numerical analysis, we normalize the sales horizon to $T=1$. Our analysis compares the following parameterized families of pricing schedules, illustrated in Figure \ref{fig:pricing_schedules}:
\begin{itemize}
    \item \textbf{Constant schedule}: The first canonical policy, defined by $p(t)=c$, where the constant price level $c$ is the single parameter.
    
    \item \textbf{Linear schedule}: The second canonical policy derived from our theory, defined by $p(t)=\max\{1-mt,0\}$, where the discount rate $m$ is the single parameter.

    \item \textbf{Polynomial schedule}:  A flexible benchmark policy defined by $p(t) = 1-(1-r)t^\alpha$. This two-parameter family allows us to explore a wide range of convex and concave price paths by varying both the terminal price $r$ and the curvature exponent $\alpha$.
    
    \item \textbf{Quasi-auction schedule}: 
    A more structured benchmark policy designed to mimic a closing auction, defined by $p(t)=1-(1-r^\star)t^\alpha$.
    In this family, the terminal price is fixed to the revenue-maximizing reservation price, $r^\star=(\lambda T -1)/(2\lambda T)$, leaving the exponent $\alpha \geq 1$ as the sole parameter.
    This allows us to specifically test the effect of concentrating the price drop near the deadline, much like a closing auction.    
\end{itemize}

For each specific schedule generated by a set of parameters, the calculation process is as follows.
First, we numerically solve the ordinary differential equation \eqref{eq:w_ode} to construct the corresponding equilibrium threshold function, $w(v)$.
While an equilibrium is guaranteed to exist for the constant and linear schedules, this is not the case for arbitrary polynomial schedules.
Therefore, for the polynomial and quasi-auction families, we also numerically verify that the sufficient conditions in Theorem~\ref{thm:VBT_suff} are met, confirming the candidate function induces a stable equilibrium.
With the equilibrium function established, we then compute the seller's expected revenue $\bar{r}_s$ and expected waiting time $\bar{\tau}_s$ by numerically evaluating the integral formulas presented in Proposition~\ref{prop:VBT_seller}.
Each resulting $(\bar{r}_s, \bar{\tau}_s)$ pair constitutes a single point, and by varying the schedule parameters, we trace out the efficient frontier for each family.

\begin{figure}
    \centering
    \begin{subfigure}[b]{0.48\linewidth}
        \includegraphics[width=\linewidth]{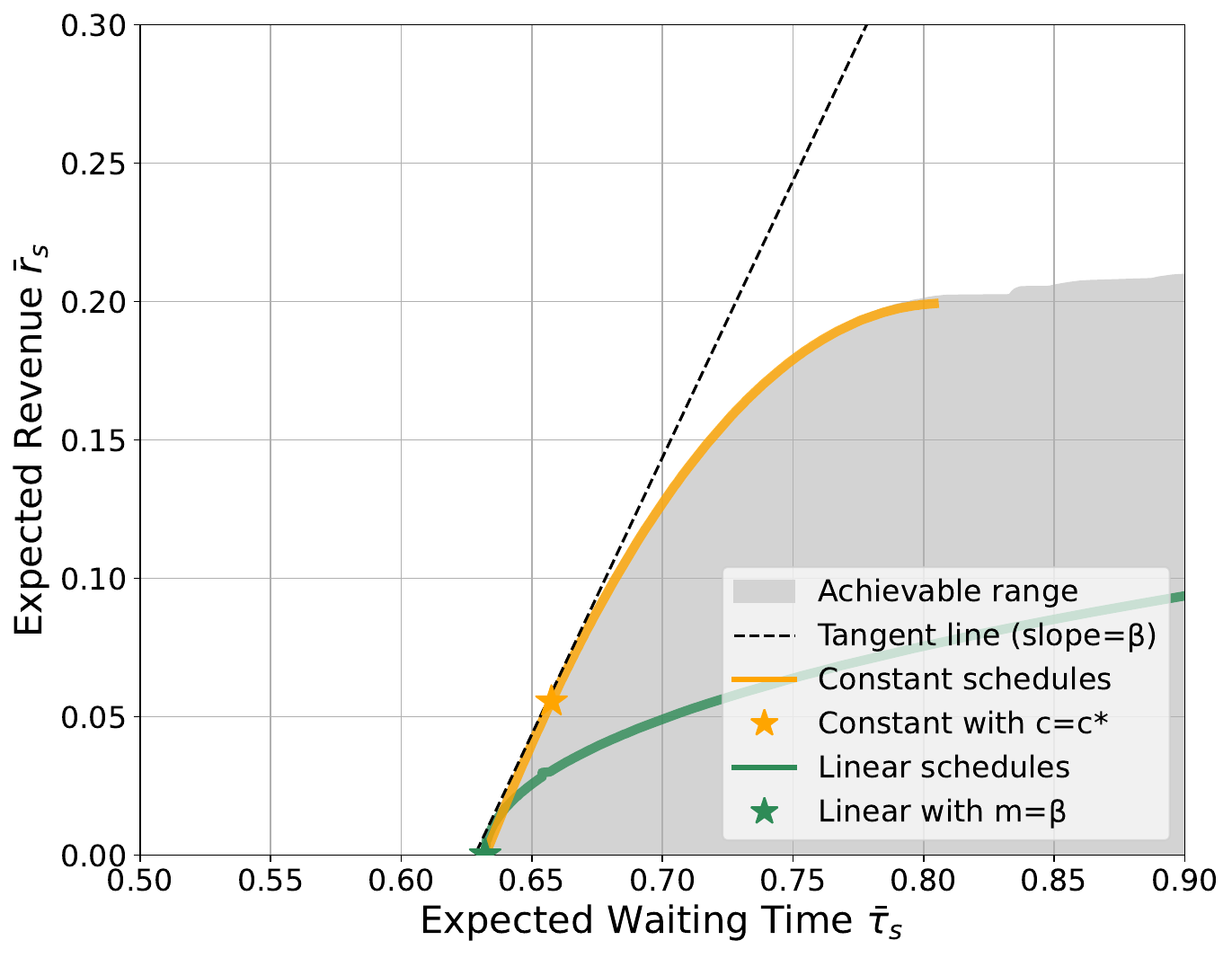}
        \captionsetup{justification=centering}
        \caption{
            Thin market ($\lambda = 1, \beta = 2$)
        }
    \label{fig:thin}
    \end{subfigure}
    \hfill
    \begin{subfigure}[b]{0.48\linewidth}
        \includegraphics[width=\linewidth]{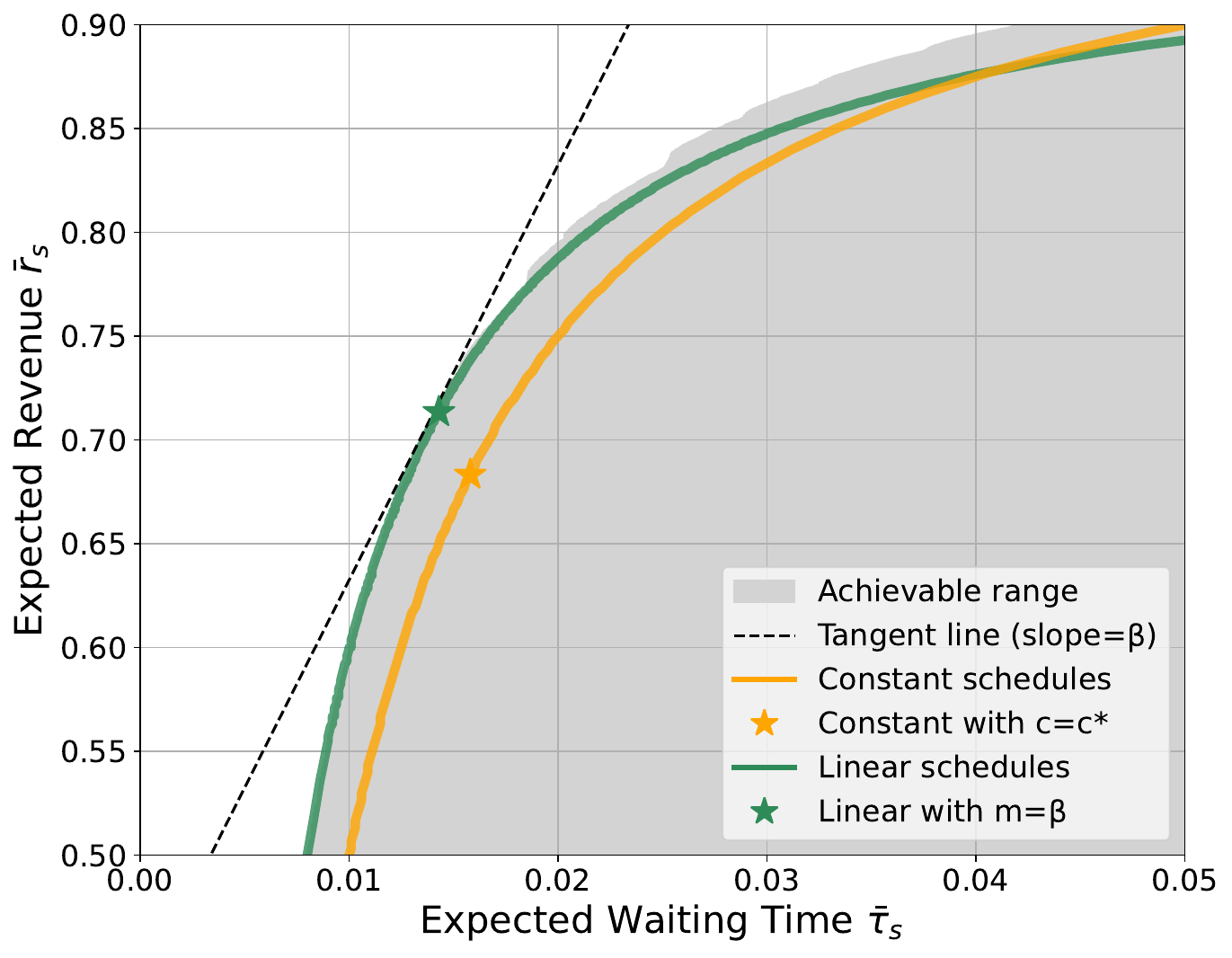}
        \captionsetup{justification=centering}
        \caption{
            Thick market ($\lambda = 200, \beta=20$)
        }        
    \label{fig:thick}
    \end{subfigure}
    \caption{Comparison of efficient frontiers and theoretical benchmarks for thick and thin market regimes.}
    \label{fig:thick_vs_thin}
\end{figure}

\paragraph{Validation of theoretical benchmarks.}
We now validate the theoretical benchmarks from Theorems \ref{thm:thin_market} and \ref{thm:thick_market} using the efficient frontier analysis. To identify the seller's optimal point in the plots, we recall their objective is to maximize $U^s = \bar{r}_s - \beta \bar{\tau}_s$, which is graphically equivalent to finding the point on the frontier tangent to a line with slope $\beta$.
In Figure~\ref{fig:thick_vs_thin}, we plot this tangent line for a seller with $\beta$ as a dashed line.
Therefore, a policy whose performance lies on this dashed line is optimal for this seller, and a policy lying close to it is near-optimal.

The results show a strong alignment between our theoretical benchmarks and this graphical optimum.
In the thin market (Figure~\ref{fig:thin}, $\lambda=1, \beta=2$), the constant pricing schedule's frontier is tangent to the line, and its theoretical benchmark (yellow star) lies on this line, validating its optimality in this regime.
Conversely, in the thick market (Figure~\ref{fig:thick}, $\lambda =200, \beta=20$), the theoretical benchmark for the linear schedule (green star) lies almost perfectly on the dashed tangent line, confirming it is near-optimal.

These results provide strong numerical support for our main theoretical findings.
For the thin market, the dominance of the constant schedule validates the ``preemptive market'' assumption as an effective approximation when competition is negligible.
Similarly, for the thick market, the close alignment of the results confirms that the ``complete impatience'' assumption is a highly accurate approximation when the arrival rate is high.
This demonstrates that our simplified analytical models effectively capture the core dynamics of the full strategic environment in these extreme regimes.

\begin{figure}
    \centering
    \begin{subfigure}[b]{0.32\linewidth}
        \includegraphics[width=\linewidth]{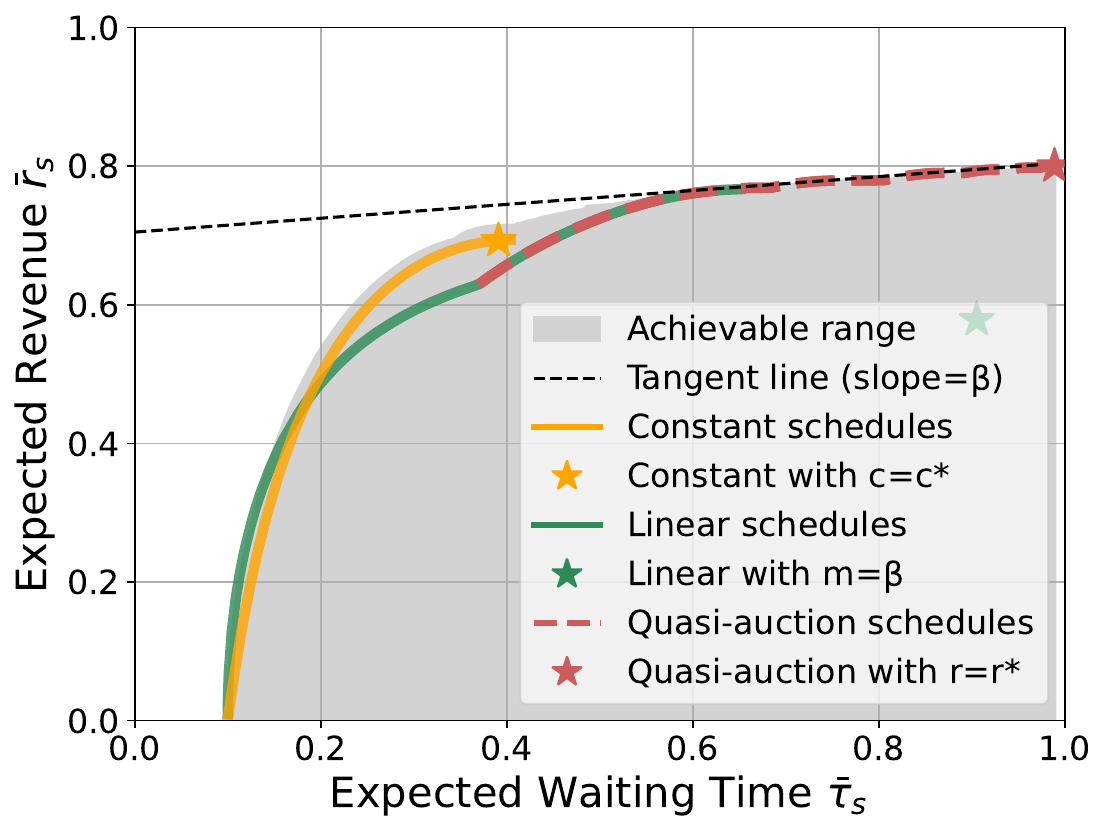}
        \captionsetup{justification=centering}
        \caption{$\lambda = 10, \beta = 0.1$}
        \label{timeinsensitive1}
    \end{subfigure}
    \hfill
    \begin{subfigure}[b]{0.32\linewidth}
        \includegraphics[width=\linewidth]{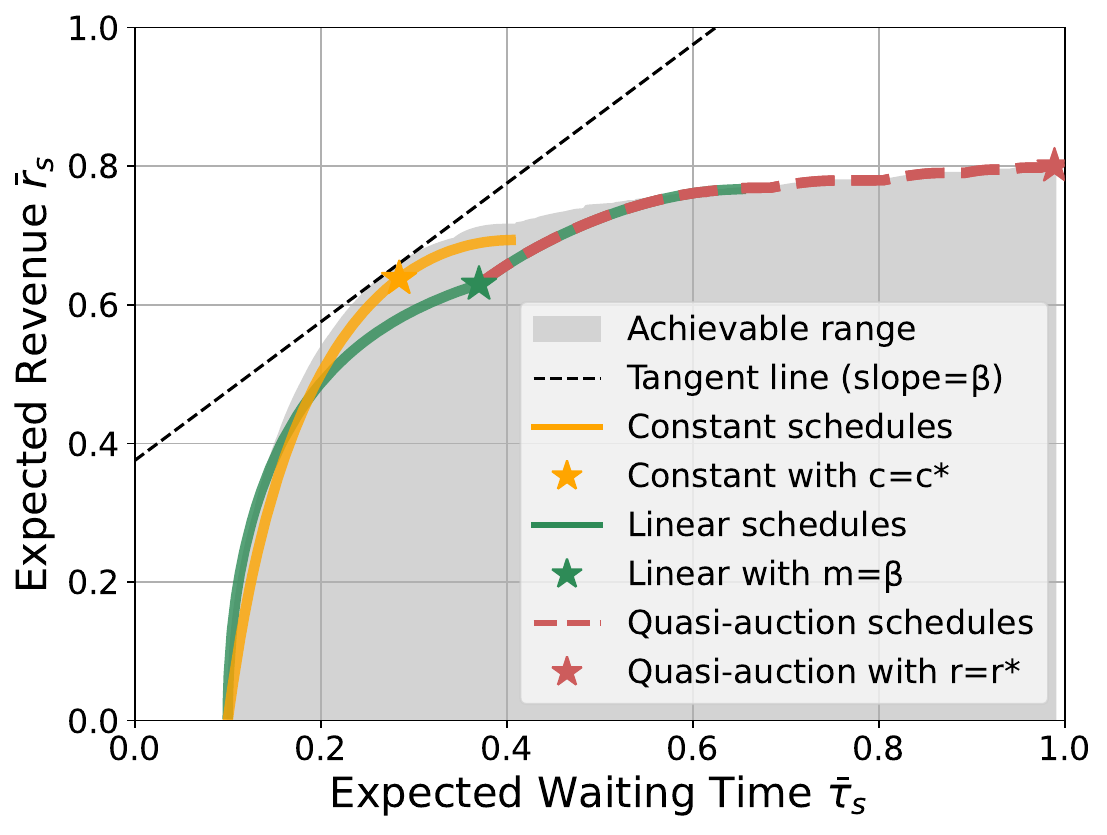}
        \captionsetup{justification=centering}
        \caption{$\lambda = 10, \beta = 1$}   
        \label{moderate1}
    \end{subfigure}
    \hfill
    \begin{subfigure}[b]{0.32\linewidth}
        \includegraphics[width=\linewidth]{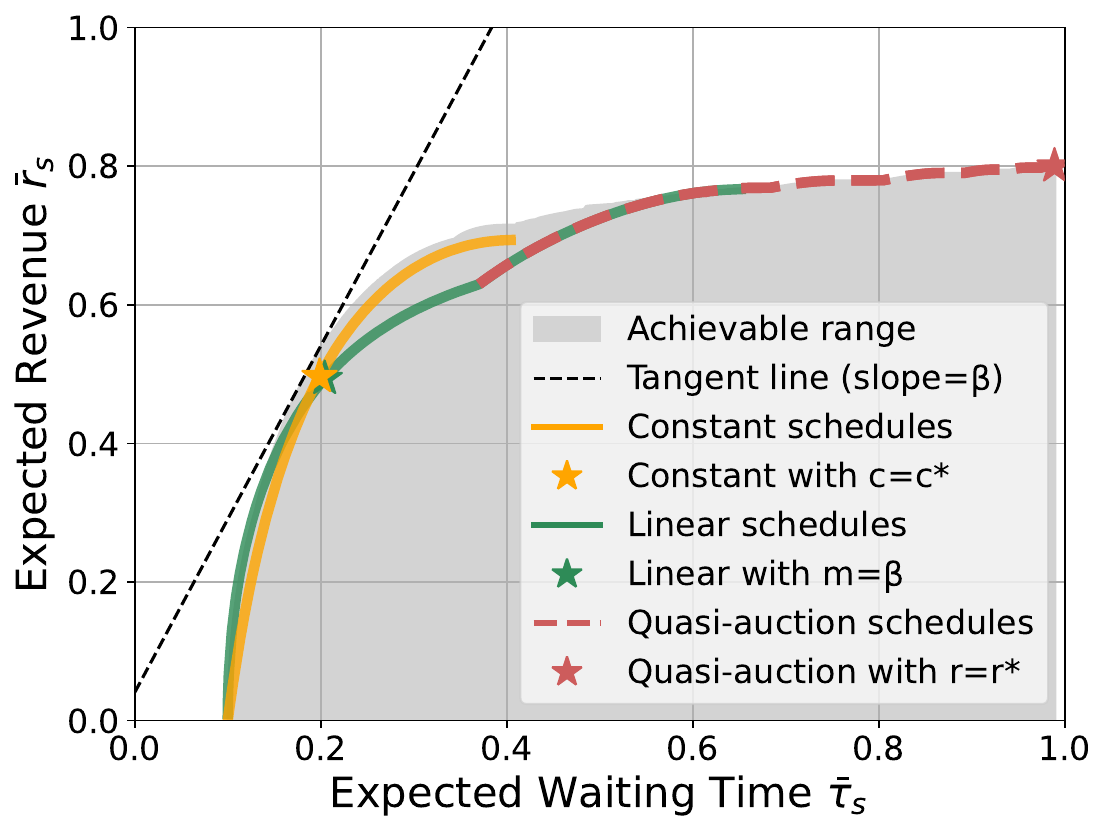}
        \captionsetup{justification=centering}
        \caption{$\lambda = 10, \beta = 2.5$}
        \label{timesensitive1}
    \end{subfigure}

    \vskip\baselineskip %

    \begin{subfigure}[b]{0.32\linewidth}
        \includegraphics[width=\linewidth]{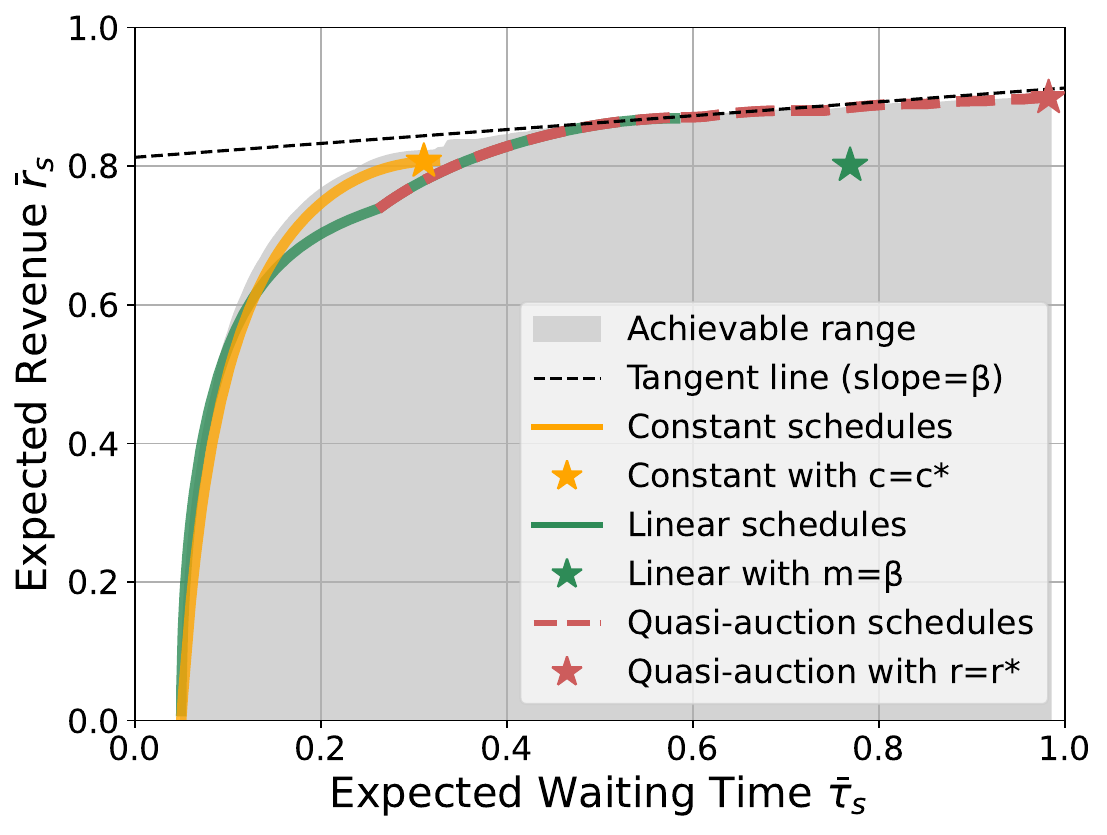}
        \captionsetup{justification=centering}
        \caption{$\lambda = 20, \beta = 0.1$}
        \label{timeinsensitive2}
    \end{subfigure}
    \hfill
    \begin{subfigure}[b]{0.32\linewidth}
        \includegraphics[width=\linewidth]{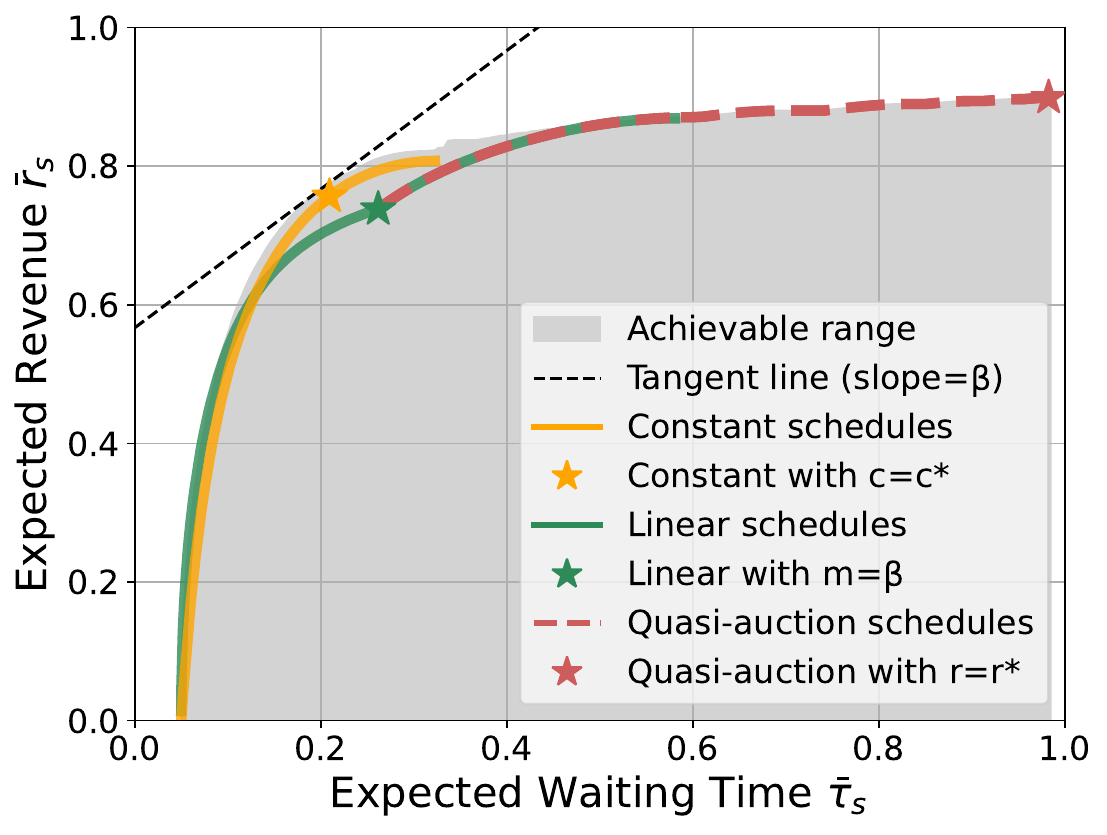}
        \captionsetup{justification=centering}
        \caption{$\lambda = 20, \beta = 1$}
        \label{moderate2}
    \end{subfigure}
    \hfill
    \begin{subfigure}[b]{0.32\linewidth}
        \includegraphics[width=\linewidth]{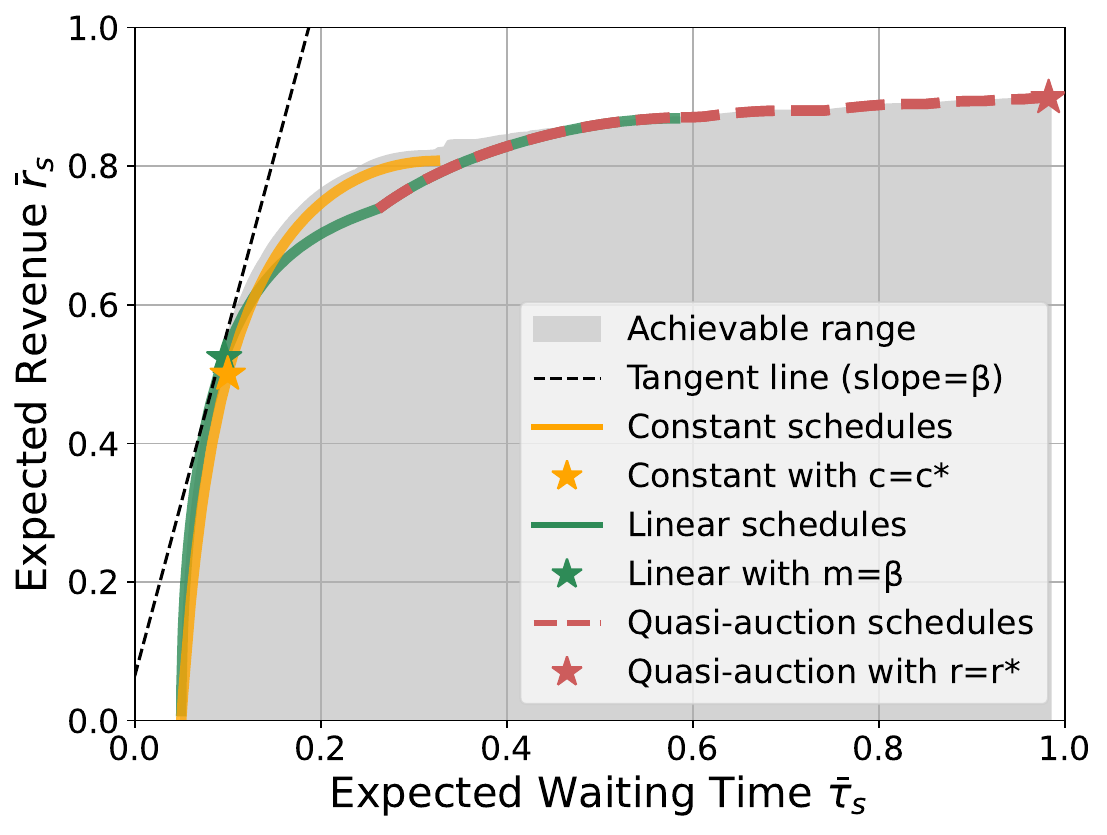}
        \captionsetup{justification=centering}
        \caption{$\lambda = 20, \beta = 5$}
        \label{timesensitive2}
    \end{subfigure}

    \caption{Comparison of efficient frontiers and theoretical benchmarks in moderate markets ($\lambda=10, 20$) across seller time sensitivity ($\beta = 0.1, 1, 2.5, 5$).}
    \label{fig:time-sensitivity}
\end{figure}

\paragraph{Comparison across levels of seller's time sensitivity ($\beta$).}
While our theoretical results provide near-optimal policies for extreme market regimes, they do not fully characterize how the optimal schedule changes with the seller's own time sensitivity, $\beta$.
We leverage our numerical efficient frontier analysis to explore this relationship in detail, with the results shown in Figure~\ref{fig:time-sensitivity}.

The analysis reveals a clear pattern.
For sellers with low sensitivity ($\beta=0.1$), the optimal schedule lies on the quasi-auction frontier.
For those with moderate sensitivity ($\beta=1$), the optimum is found on the constant schedule frontier.
For high sensitivity ($\beta \geq 2.5$), the linear schedule frontier performs best, though the constant schedule remains a strong competitor, especially in thinner markets.

These observations lead to clear strategic recommendations.
\emph{For a time-insensitive seller, a quasi-auction schedule is the most attractive solution.}
With little pressure to sell quickly, the seller can afford to maintain a high price for most of the horizon, aggregating future demand. A sharp, terminal price drop then triggers a late burst of competition, maximizing revenue.

\emph{For a seller with moderate/high time sensitivity, the constant schedule is a robust, near-optimal choice.}
The key decision for the seller is not the shape of the price curve, but rather optimizing the single price level, $c$.
This simple policy provides the best trade-off between revenue and waiting time for a moderately patient seller.

Finally, \emph{for a highly time-sensitive seller, especially in a thick market, the linear schedule is an effective and simple policy.}
A key advantage is its simplicity: the optimal discount rate is determined solely by the seller's own time sensitivity $\beta$, not by complex market conditions like arrival rate $\lambda$.
This allows an impatient seller to implement a near-optimal strategy without needing to optimize for market thickness or the time horizon.

\section{Conclusion} \label{sec:conclusion}
This research advances the dynamic pricing literature by presenting a framework that integrates time-sensitive single seller with strategic buyers with stochastic arrival. 
The model is formulated as a continuous-time, single-leader multi-follower game. 
Positioned at the intersection of dynamic pricing, optimal stopping theory, and auction design, offers a multidisciplinary perspective on pricing policy.

We address the dynamic pricing problem through a two-stage framework. 
First, we characterize the equilibrium of strategic buyers with heterogeneous arrival times and private valuations. 
The dual dependence on type complicates the equilibrium structure, as earlier arrival confers a strategic advantage.
To address this complexity, we introduce the Value-Based Threshold (VBT) strategy, characterized by a threshold time function. 
This reduces the buyer’s decision to a comparison between arrival time and threshold time, thereby simplifying the equilibrium analysis in markets with stochastic arrivals.
The VBT framework enhances both tractability and intuition, offering a systematic approach to capturing the interplay between price dynamics and arrival times. 
This contributes to the broader literature on optimal pricing in dynamic markets.

Furthermore, we propose near-optimal pricing strategies for two extreme market regimes ---thick and thin markets---demonstrating how the seller's optimal policy adapts to varying buyer arrival rates.
In thick markets, where buyers arrive frequently, strategic delay offers no incentive, thus buyer behavior becomes effectively myopic. 
Assuming no delay, we show that the optimal pricing policy follows a linear discounting, with a slope determined by the seller’s time sensitivity coefficient. 
Simulation results confirm that the optimal pricing schedule aligns closely with this theoretical result.

In contrast, thin markets with infrequent buyer arrivals are analyzed under the extreme assumption that the first-arriving buyer holds a preemptive right to purchase. 
Under this setting, we establish that the seller’s optimal pricing policy is constant. 
As the arrival rate decreases, the market approaches this preemptive structure. 
Simulations further demonstrate that, at sufficiently low arrival rates, the near-optimal pricing policy converges to a constant pricing schedule. 
This insight aligns with empirical findings; according to \citet{wang2008auction}, fixed pricing maximizes seller utility when the number of prospective buyers is small, as observed on online market platforms like eBay. 

Despite its contributions, a primary limitation of this research is that it assumes a continuous pricing schedule.
Consequently, our framework does not formally cover discontinuous price drops, including the ones with a closing auction.
While our quasi-auction schedule can approximate such a mechanism, it does not model it strictly. 
More generally, any schedule with multiple distinct price drops (or continuous but extremely steep price declines) could create a scenario where a buyer's optimal strategy depends on both their valuation and arrival time.
This would violate the core decoupling property of the VBT framework, potentially rendering our equilibrium analysis invalid.
Characterizing the optimal policy for discontinuous schedules or hybrid buyer strategies remains a significant analytical challenge and a promising direction for future research.

A further limitation is that while our framework fully characterizes the buyer's equilibrium for a given price schedule, it does not provide a general analytical characterization of the seller's optimal price schedule.
Although a numerical solution for the optimal schedule is possible in principle (e.g., by computing the sensitivity of the ODE solution with respect to the pricing schedule), a more fruitful direction for future research is to better characterize the properties of the optimal price path or to establish more formal bounds on the near-optimality of the simple canonical schedules we propose.
Our numerical analysis offers a hint in this direction: the constant price schedule, in particular, performs remarkably well across a wide range of market conditions, not just in the extremely thin regime.
Understanding the theoretical reasons for the robustness of such simple policies remains a valuable and challenging question.

\bibliographystyle{apalike}  %
\bibliography{references}

\appendix %
\renewcommand{\theequation}{\arabic{equation}}   
\setcounter{equation}{0}

\section{Proofs} \label{app:proof}

\subsection{Proof of Proposition~\ref{prop:VBT_buyer}}
\begin{proof}
    Fix $t \in \Tc = [0,T]$. 
    The item remains unsold until time $t$ if no buyer with $v > w^{-1}(t)$ arrives before $t$. 
    Since $v \sim \text{Uniform}[0,1]$, the arrival of such buyers follows a Poisson process with rate $\lambda (1 - w^{-1}(t))$. 
    Thus, the probability that the item is unsold at $t$ is
    \begin{equation} \label{unsold_prob}
        \Pr( \tau^* > t ) = e^{-\lambda t\big(1-w^{-1}(t)\big)}, \quad \forall t \in \Tc.
    \end{equation}
    Then the interim utility of a buyer is given by the product of his payoff and the probability that the item remains unsold untill his target purchase time~ $\tau$ as follows:
    \begin{align}
    \Pi^b(\tau; v, \alpha, \sigma_w, p) 
    &= \Pi^b(\tau; v, 0, \sigma_w, p) \notag \\
    &= (v - p(\tau)) e^{-\lambda  \tau (1 - w^{-1}(\tau))}, \quad \forall \tau \in [\alpha,T]. 
    \end{align}
    \qedhere
\end{proof}

\subsection{Proof of Proposition~\ref{prop:VBT_seller}}
\begin{proof}
    Assume $N \sim \text{Poisson}(\lambda T)$ buyers arrive over the horizon $\Tc = [0,T]$. 
    Under the Value–Arrival Separable (VAS) strategy, the buyer $i$'s target purchase time is defined as
    \[
    \tau_i := \max\{\alpha_i, w(v_i)\},
    \]
    Define the earliest target purchase time and seller's waiting time as
    \[
    	\tau^* = \min_{i \in [N]} \tau_i, \quad \tau_s = \min\{\tau^*, T\}.
    \]
    Therefore, the seller's expected waiting time $\bar{\tau}_s$ is given by 
    \begin{equation}
    	\bar{\tau}_s = \mathbb{E}[\tau_s] = \int_0^T \Pr[\tau_s > t] \dt.
    \end{equation}

    Since $\tau_s > t$ if and only if $\tau_i > t$ for all $i \in [n]$, and assuming that $\{\tau_i\}_{i=1}^n$ are i.i.d., it follows that 
    \begin{equation}
        \Pr[\tau_s > t \mid N = n] = \prod_{i=1}^n \Pr[\tau_i > t] = \left( \Pr[\tau_i > t] \right)^n. \label{eq:prob_tau_s_cond}
    \end{equation}
    Combining the conditional probability with the distribution of $N$, we obtain
    \begin{align}
        \Pr[\tau_s > t]
        &= \sum_{n=0}^\infty \left( \Pr[\tau_i > t] \right)^n \cdot \Pr[N = n] \notag \\
        &= \sum_{n=0}^\infty \left( \Pr[\tau_i > t] \right)^n \cdot \frac{(\lambda T)^n e^{-\lambda T}}{n!} \notag \\
        &= \mathbb{E}_N \left[ \left( \Pr[\tau_i > t] \right)^N \right].
        \label{eq:prob_tau_s}
    \end{align}

    We now compute the probability that a buyer’s target purchase time $\tau$ exceeds $t$,
    \begin{align}
        \Pr[\tau_i > t] 
        &= \Pr[\max\{\alpha_i, w(v_i)\} > t] \\
        &= \Pr[\alpha_i > t \lor w(v_i) > t] \\
        &= 1 - \Pr[\alpha_i \leq t,\, w(v_i) \leq t] \label{eq:prob_phi},
    \end{align}
    Since $\alpha_i$ and $v_i$ are independent, we have
    \begin{align}
        \Pr[\alpha_i \leq t,\, w(v_i) \leq t] 
        &= \Pr[\alpha_i \leq t] \cdot \Pr[w(v_i) \leq t] \\
        &= \frac{t}{T} \cdot \Pr[v_i \leq w^{-1}(t)] \\
        &= \frac{t}{T} \cdot (1 - w^{-1}(t)),
    \end{align}
    by the regularity condition that the threshold time function $w$.
    Substituting into \eqref{eq:prob_phi}, we obtain
    \begin{align}
        \Pr[\tau_i > t] 
        &= 1 - \frac{t}{T}(1 - w^{-1}(t)) \notag \\
        &= 1 - \frac{t}{T} + \frac{t}{T} w^{-1}(t). \label{eq:phi}
    \end{align}
    Define
    \[
    \phi(t) := 1 - \frac{t}{T} + \frac{t}{T} w^{-1}(t).
    \]
    Then, from \eqref{eq:prob_tau_s_cond}  the conditional probability becomes
    \begin{align}
	    \Pr[\tau_s > t \mid N = n] = \phi^n(t).
    \end{align}
    Taking expectation over $N \sim \text{Poisson}(\lambda T)$, and substituting the result into \eqref{eq:prob_tau_s}, we have
    \begin{align}
        \Pr[\tau_s > t] 
        &= \sum_{n=0}^\infty \phi^n(t) \cdot \Pr[N = n] \\
        &= \sum_{n=0}^\infty \phi^n(t) \cdot \frac{(\lambda T)^n e^{-\lambda T}}{n!} \\
        &= e^{-\lambda T} \sum_{n=0}^\infty \frac{(\lambda T \cdot \phi(t))^n}{n!} \\
        &= e^{-\lambda T (1 - \phi(t))} \\
        &= e^{-\lambda t(1 - w^{-1}(t))}.   \label{eq:prob_tau_s_final}
    \end{align} 
    The fourth equality follows from the Taylor expansion of the exponential function,  \( e^x = \sum_{n=0}^\infty \frac{x^n}{n!} \).
    Therefore, the seller's expected waiting time is
	\[
    	\bar{\tau}_s = \int_0^T e^{-\lambda t(1 - w^{-1}(t))} \dt.
    \]

    Let \( r_s := p(\tau^*) \cdot \mathbb{I}\{\tau^* < \infty\} \) denote the seller’s realized revenue.
    The expected revenue is then given by
    \begin{equation}
        \bar{r}_s = \mathbb{E}[r_s] = \mathbb{E}[p(\tau^*) \cdot \mathbb{I}\{\tau^* < \infty \}].
    \end{equation}
    Similar to the expected waiting time, the expected revenue is given by  
    \begin{equation}
    \bar{r}_s = \int_0^{p(0)} \Pr[r_s> r] \,\dr
    \end{equation}
    To evaluate this expression, we partition the integral over the revenue variable \( r \) based on whether \( r \) is below or above the terminal price  \( p(T) \).  
    \begin{align}
    \bar{r}_s 
    &= \int_0^{p(T)} \Pr[r_s > r] \, \dr + \int_{p(T)}^{p(0)} \Pr[r_s > r] \, \dr \notag \\
    &= \int_0^{p(T)} \left( 1 - \Pr[r_s < r] \right) \, \dr + \int_{p(T)}^{p(0)} \left( 1 - \Pr[r_s < r] \right) \, \dr \notag \\
    &= {p(0)} - \int_0^{p(T)} \Pr[r_s < r] \, \dr - \int_{p(T)}^{p(0)} \Pr[r_s < r] \,\dr \label{eq:revenue_partition}
    \end{align}

    We first compute \( \Pr[r_s < r] \) for \( r \in [0, p(T)] \).  
    It follows that \( r_s < r \) only if no buyer arrives with a valuation high enough to induce purchase within the horizon, so we have  
    \begin{equation}
    \Pr[r_s < r] = \Pr[\tau^* = \infty] = e^{-\lambda T(1 - p(T))}. \label{eq:rs_less_than_r_case1}
    \end{equation}

    Next, consider \( r \in [p(T), 1] \).  
    On the event \( \tau^* < \infty \), we have \( r_s = p(\tau^*) \), and thus by invertibility of \( p \),  
    \[
    \tau^* = p^{-1}(r_s).
    \]
    Therefore, for \( r \geq p(T) \), the event \( r_s < r \) is equivalent to \( \tau^* > p^{-1}(r) \), and we obtain
    \begin{align}
    \Pr[r_s < r] &= \Pr[\tau^* > p^{-1}(r)] \notag \\
    &= e^{-\lambda p^{-1}(r)(1 - w^{-1}(p^{-1}(r)))}. \label{eq:rs_less_than_r_case2}
    \end{align}
    Substituting \eqref{eq:rs_less_than_r_case1} and \eqref{eq:rs_less_than_r_case2} into \eqref{eq:revenue_partition}, we obtain
    \begin{align}
    \bar{r}_s 
    &= {p(0)} - p(T) \cdot e^{-\lambda T (1 - p(T))} - \int_{p(T)}^{p(0)} e^{-\lambda p^{-1}(r) (1 - w^{-1}(p^{-1}(r)))} \,\dr \label{eq:revenue_final1}
    \end{align}

    Now, make the change of variables \( r = p(t) \), which implies \( dr = p'(t) \dt \).  
    Changing variables in the second integral of \eqref{eq:revenue_final1} yields:
    \begin{align}
    \bar{r}_s 
    &= {p(0)} - p(T) \cdot e^{-\lambda T (1 - p(T))} + \int_0^T e^{-\lambda t(1 - w^{-1}(t))} \cdot p'(t) \, \dt. \label{eq:revenue_final2}
    \end{align}
    \qedhere
\end{proof}

\subsection{Proof of Theorem~\ref{thm:VBT_necc}}
\begin{proof}
    Since $w$ induces an equilibrium, it must satisfy the buyer’s best response condition for all $v \in \Vc = [0,1]$ and $\alpha \in \Tc = [0,T]$. 
    In particular, when $\alpha = 0$,
    \begin{align*}
    \sigma_w(v,0) &= \max\{w(v), 0\} = w(v) \notag \\
    &= \inf \argmax_{\tau \in [0,T] \cup \{\infty\}} \Pi^b(\tau; v, 0, \sigma_w, p), \quad \forall v \in [0,1]. 
    \end{align*}
    We consider each of the four cases as follows.
    \begin{enumerate}[(i)]
    \item Fix $v \in [0, p(T))$. Since $p(\cdot)$ is non-increasing,
        \[
        	p(\tau) \geq p(T) > v, \quad \forall \tau \in [0,T],
        \]
	    which implies
        \begin{align*}
        	\Pi^b(\tau; v, 0, \sigma_w, p) &= (v - p(\tau)) e^{-\lambda  \tau (1 - w^{-1}(\tau))} < 0, \\
        	\Pi^b(\infty; v, 0, \sigma_w, p) &= 0.
        \end{align*}
        Therefore,
        \[
        	\argmax_{\tau \in [0,T] \cup \{\infty\}} \Pi^b(\tau; v, 0, \sigma_w, p) = \{\infty\} ,
        \]
        and, consequently,
        \[
		     w(v) = \infty.
        \]
        That is, buyers with values below the minimum price $p(T)$ never purchase.
    
    \item Let $v = p(T)$ and define
        \begin{align*}
        	\tau^* := \inf\{t \in [0,T] : p(t) = p(T)\}.
        \end{align*}
        By continuity and monotonicity of $p(\cdot)$,
        \[
        	p(\tau) > p(\tau^*) = v, \quad \forall \tau \in [0,\tau^*), 
        \]
        which implies
        \begin{align*}
        	\Pi^b(\tau; v, 0, \sigma_w, p) &< 0, \quad \forall \tau \in [0,\tau^*), \\
	        \Pi^b(\tau; v, 0, \sigma_w, p) &= 0, \quad \forall \tau \in [\tau^*, T] \cup \{\infty\}.
        \end{align*}
        Thus,
        \[
        	\argmax_{\tau \in [0,T] \cup \{\infty\}} \Pi^b(\tau; v, 0, \sigma_w, p) = [\tau^*, T] \cup \{\infty\}
        \]
        By the characterization of the best response in \eqref{eq:equilibrium}, 
        \[
        	w(v) = \tau^*.
        \]
        In this case, the buyer purchases as soon as the price drops to $p(T)$.
        
    \item For $v \in [p(T),1)$, applying interior extremum theorem, then the first-order condition gives
        \[
        	\left. \frac{\dd}{\dd\tau} \Pi^b(\tau; v, 0, \sigma_w, p) \right|_{\tau = w(v)} = 0.
        \]
        From Proposition~\ref{prop:VBT_buyer},
        \begin{align*}
        	0 
			&= \left. \frac{\dd}{\dd\tau} \left[ (v - p(\tau)) e^{-\lambda \tau (1 - w^{-1}(\tau))} \right] \right|_{\tau = w(v)} \notag \\
	        &= e^{-\lambda w(v)(1 - v)} \left[ \lambda \left(-1 + v + \frac{w(v)}{w_+'(v)} \right)(v - p(w(v))) - p'(w(v)) \right],
        \end{align*}
        which yields the identity:
        \[
        	\lambda \left(-1 + v + \frac{w(v)}{w_+'(v)} \right)(v - p(w(v))) = p'(w(v)).
        \]
        
    \item We verify $p(w(v)) \leq v$ for $v \in [p(T),1]$. Suppose not, i.e., $p(w(v)) > v$. Then,
	    \begin{align*}
    		\Pi^b(w(v); v, 0, \sigma_w, p) &= (v - p(w(v))) e^{-\lambda   w(v)(1 - v)} < 0, \\
	    	\Pi^b(\infty; v, 0, \sigma_w, p) &= 0,
	    \end{align*}
    	which contradicts economic rationality. 
		Hence, $p(w(v)) \leq v$.
    \end{enumerate}
    \qedhere
\end{proof}

\subsection{Proof of Theorem~\ref{thm:VBT_suff}}
\begin{proof}
    Let $\sigma_w$ be the VAS strategy induced by the given threshold function $w$. We aim to prove:
    \begin{align}
        \max\{w(v), \alpha\} 
        = \inf \argmax_{\tau \in [\alpha,T] \cup \{\infty\}} \Pi^b(\tau; v, \alpha, \sigma, p), 
        \quad \forall v \in \Vc = [0,1], 
        \quad \forall \alpha \in [0,T]. \label{eq:goal}
    \end{align}
    By Proposition~\ref{prop:VBT_buyer}, for all $\tau \in [\alpha,T]$,
    \begin{align}
        \Pi^b(\tau; v, \alpha, \sigma_w, p) 
        = \Pi^b(\tau; v, 0, \sigma_w, p) 
        = (v - p(\tau)) e^{-\lambda \tau (1 - w^{-1}(\tau))}. \label{eq:prop}
    \end{align}
    Hence, it suffices to verify:
    \begin{align}
        \max\{w(v), \alpha\} 
        = \inf \argmax_{\tau \in [\alpha,T] \cup \{\infty\}} \Pi^b(\tau; v, 0, \sigma_w, p), \quad \forall v \in [0,1],
        \quad \forall \alpha \in [0,T]. \label{eq:reduced-goal}
    \end{align}
    We consider two cases for the $v$ values. 

    Suppose $v \in [0, p(T))$. 
    Since $p(\cdot)$ is non-increasing, we have
        \begin{align*}
	        p(\tau) \geq p(T) > v, \quad \forall \tau \in [\alpha, T],
        \end{align*}
        and, therefore,
        \begin{align*}
            \Pi^b(\tau; v, 0, \sigma, p) 
            &= (v - p(\tau)) e^{-\lambda   \tau (1 - w^{-1}(\tau))} < 0, \quad \forall \tau \in [\alpha,T], \\
            \Pi^b(\infty; v, 0, \sigma, p) &= 0.
        \end{align*}
        Thus,
        \begin{align*}
            \argmax_{\tau \in [\alpha,T] \cup \{\infty\}} \Pi^b(\tau; v, 0, \sigma, p) = \{\infty\},
        \end{align*}
        which implies
        \begin{align*}
            \max\{w(v), \alpha\} = \infty = \inf \argmax_{\tau \in [\alpha,T] \cup \{\infty\}} \Pi^b(\tau; v, 0, \sigma, p).
        \end{align*}
        This holds for all $\alpha \in [0,T]$.

        Consider now the case where $v \in [p(T), 1)$. 
        Under the condition of this theorem, the set of global maximizers of 
        $\Pi^b(\cdot; v, 0, \sigma_w, p)$ on $[0,T]$ is connected. 
        By the definition of the best response \eqref{eq:equilibrium}, we have
        \begin{align}
              w(v) \;=\; \inf \argmax_{\tau \in [0,T]} \Pi^b(\tau; v, 0, \sigma_w, p), \qquad \forall v \in [p(T),1).
        \end{align}
    
        We now examine two subcases based on the relation between $w(v)$ and $\alpha$.
        First, if $\alpha \leq w(v)$, then $w(v) \in [\alpha,T]$ and it follows: 
        \begin{align}
            \inf \argmax_{\tau \in [\alpha,T] \cup \{\infty\}} \Pi^b(\tau; v, 0, \sigma, p) = w(v) = \max\{w(v), \alpha\}.
        \end{align}
        for all $v \in [p(T),1)$ and $\alpha \in [0, w(v)]$, where the last equality is by definition of $\sigma_w$.

        On the other hand, if $\alpha > w(v)$, then 
        $\Pi^b(\cdot; v, 0, \sigma_w, p)$ is non-increasing on $[\alpha,T]$, 
        since its global maximizer set is a connected subset of $[0,T]$ with infimum $w(v)$. 
        Thus, 
        \begin{align}
            \inf \argmax_{\tau \in [\alpha,T] \cup \{\infty\}} \Pi^b(\tau; v, 0, \sigma_w, p)  = \alpha = \max\{w(v), \alpha\},
        \end{align}
        for all $v \in [p(T),1)$ and $\alpha \in (w(v),T]$, where the last equality is by definition of $\sigma_w$.
    
    \qedhere
\end{proof}
    
\subsection{Proof of Theorem~\ref{thm:thin_market}}
\begin{proof}
    Define the terminal price \( c := p(T) \in (0,1) \). 
    In a preemptive market, the first-arrived buyer occupy the market without the need to purchase the item immediately.
    Therefore, the first-arrived buyer who has a private valuation larger than or equal to the terminal price \( c \) will purchase the item at time \( t \) when the price equals \( c \).
    \begin{equation*}
        \tau^* = \begin{cases}
            \infty & \text{if } v < c, \\
            \inf\{t \in [\alpha, T] : p(t) = c\} & \text{if } v \geq c,
        \end{cases}
    \end{equation*}
    Note that the seller's expected revenue and waiting time are given by
    \begin{align*}
        \bar{r}_s &= \mathbb{E}[p(\tau^*) \cdot \mathbb{I}\{\tau^* < \infty\}], \quad \bar{\tau}_s = \mathbb{E}[\min\{\tau^*, T\}].
    \end{align*}
    Now calculate the seller's expected revenue first. 
    \begin{align*}
        \bar{r}_s &= (1-e^{-\lambda T(1-c)})\cdot c + e^{-\lambda T(1-c)}\cdot0\\
        &= (1-e^{-\lambda T(1-c)})c.
    \end{align*}

    Next, we compute the seller's expected waiting time. 
    The expected waiting time is given by
    \[
    \bar{\tau}_s = \mathbb{E}[\min\{\tau^*, T\}].
    \]
    We consider the conditional expectation depending on arrival.
    \begin{align*}
        \bar{\tau}_s 
        &= \mathbb{E}_\alpha[\tau_c(\alpha)] \cdot (1 - e^{-\lambda T((1 - c))}) + T \cdot \left( 1 - (1 - e^{-\lambda T(1 - c)}) \right) \\
        &= \mathbb{E}_\alpha[\tau_c(\alpha)] \cdot (1 - e^{-\lambda T((1 - c))}) + T e^{-\lambda T(1 - c)}.
    \end{align*}
    where
    \[
    \tau_c(\alpha) := \inf\{t \in [\alpha, T] : p(t) = c\}.
    \]

    We now characterize the optimal price schedule that maximizes the seller's expected utility.

    Note that \( \tau_c(\alpha) \geq \alpha \) for any price schedule, and equality holds if and only if \( p(t) = c \) for all \( t \in [0, T] \). Therefore,
    \[
    \mathbb{E}_\alpha[\tau_c(\alpha)] \geq \mathbb{E}_\alpha[\alpha] = \frac{T}{2},
    \]
    with equality only when \( p(t) \equiv c \). 
    Therefore, the optimal price schedule in a preemptive market is the constant price schedule \( p^*(t) = c \) for all \( t \in [0, T] \).

    Finally, we examine the comparative statics of the optimal constant price \( c^*(\lambda, \beta) \) with respect to the buyer arrival rate \( \lambda \) and the seller’s time sensitivity coefficient \( \beta \). 
    Under a constant price schedule \( p(t) \equiv c \), the seller’s expected utility \( U^s(c) \) is given by
    \begin{align*}  
        U^s(c) 
        &= c \cdot (1 - e^{-\lambda T (1 - c)}) - \beta \left( \frac{T}{2} (1 - e^{-\lambda T(1 - c)}) + T e^{-\lambda T(1 - c)} \right).
    \end{align*}
    To identify the optimal price, we differentiate \( U^s(c) \) with respect to \( c \).
    The first-order condition yields:
    \begin{align*}
        \frac{\dd U^s}{\dd c} 
        &= 1 - e^{-\lambda T (1 - c)} - \left( c + \frac{1}{2} \beta T \right) \lambda T e^{-\lambda T (1 - c)} = 0.
    \end{align*}
    Solving the above condition gives the following implicit characterization of $\hat{c}$ that satisfies the first-order condition:
    \[
    \hat{c}(\lambda, \beta, T) = 1 - \frac{1}{\lambda T} \ln\left(1 + \lambda T \left( \hat{c}(\lambda, \beta, T) + \tfrac{1}{2} \beta T \right) \right).
    \]

    Let
    \[
    F(c;\lambda,\beta,T)
    :=c-1+\frac{1}{\lambda T}
    \ln\!\bigl[\,1+\lambda T\bigl(c+\tfrac12\beta T\bigr)\bigr].
    \]
    By direct calculation
    \[
    \frac{\partial F}{\partial c}
    =1+\frac{1}{1+\lambda T(c+\tfrac12\beta T)}>0,
    \]
    Moreover,
        \[
        \lim_{c\to-\infty}F(c)=-\infty,
        \qquad
        \lim_{c\to+\infty}F(c)=+\infty,
        \]
    hence by the intermediate‐value theorem there is exactly one real root \(\hat c\in \mathbb{R}\) satisfying \(F(\hat c;\lambda,\beta,T)=0\).  
    
    Define the \emph{constrained} optimal price $c^*$ by projecting onto \([0,1]\):
    \[
    c^\star(\lambda,\beta,T)
    =\begin{cases}
    0,&\hat c<0,\\
    \hat c,&0\le\hat c\le1,\\
    1,&\hat c>1.
    \end{cases}
    \]
    Observe that under a constant schedule \(p(t)\equiv c\),
    \[
    U^s(c)=c\bigl(1-e^{-\lambda T(1-c)}\bigr)
    -\beta\Bigl[\tfrac T2(1-e^{-\lambda T(1-c)})+T\,e^{-\lambda T(1-c)}\Bigr],
    \]
    and one checks
    \[
    U'(c)=0
    \quad\Longleftrightarrow\quad
    F(c;\lambda,\beta,T)=0.
    \]
    Since \(F\) is strictly increasing in \(c\), the sign of \(U'\) flips
    exactly once at \(c=\hat c\):
    \[
    U'(c)>0\;(c<\hat c),
    \qquad
    U'(c)<0\;(c>\hat c).
    \]
    By the extreme‐value theorem on the compact interval \([0,1]\), the
    global maximizer is
    \[
    \begin{cases}
    \hat c, & \text{if }0\le\hat c\le1,\\
    0, & \text{if }\hat c<0,\\
    1, & \text{if }\hat c>1,
    \end{cases}
    \]
    which is precisely \(c^\star(\lambda,\beta,T)\).
    
    Now we compute the comparative statics of the constrained optimal price \(c^\star(\lambda,\beta,T)\) with respect to \(\lambda\), \(\beta\) and T.
    On the interior region \(0<\hat c<1\), the implicit‐function theorem applied to
    \[
    F\bigl(\hat c;\lambda,\beta,T\bigr)=0
    \]
    gives
    \[
    \frac{\partial \hat c}{\partial x}
    =-\frac{F_x}{F_c},
    \quad x\in\{\lambda,\beta,T\}.
    \]
    A direct calculation yields
    \[
    F_c(c)
    =1+\frac{1}{1+\lambda T\bigl(c+\tfrac12\beta T\bigr)}>0,
    \]
    \[
    F_\lambda(c)
    =-\frac{1}{\lambda^2 T}\,
    \ln\!\bigl[1+\lambda T\bigl(c+\tfrac12\beta T\bigr)\bigr]
    \;+\;\frac{c+\tfrac12\beta T}{\lambda\bigl(1+\lambda T(c+\tfrac12\beta T)\bigr)},
    \]
    \[
    F_\beta(c)
    =\frac{1}{\lambda T}\,
    \frac{\partial}{\partial\beta}\ln\!\bigl[1+\lambda T(c+\tfrac12\beta T)\bigr]
    =\frac{T}{2\bigl(1+\lambda T(c+\tfrac12\beta T)\bigr)}>0,
    \]
    \[
    F_T(c)
    =-\frac{1}{\lambda T^2}\,
    \ln\!\bigl[1+\lambda T(c+\tfrac12\beta T)\bigr]
    \;+\;\frac{1}{\lambda T}
    \cdot\frac{\lambda\bigl(c+\beta T\bigr)}{\,1+\lambda T(c+\tfrac12\beta T)\,}
    =\,-\frac{1-c}{T}+\frac{c+\beta T}{T\bigl(1+\lambda T(c+\tfrac12\beta T)\bigr)}.
    \]

    Since at \(c=\hat c\) the identity
    \(\ln[1+\lambda T(\hat c+\tfrac12\beta T)]=\lambda T(1-\hat c)\) holds,
    one checks
    \[
    F_\lambda(\hat c)<0,\qquad
    F_\beta(\hat c)>0,
    \]
    and under the same parameter conditions guaranteeing \(0<\hat c<1\) one also shows
    \[
    F_T(\hat c)<0.
    \]
    Therefore
    \[
    \frac{\partial \hat c}{\partial \lambda}
    =-\frac{F_\lambda}{F_c}>0,
    \quad
    \frac{\partial \hat c}{\partial \beta}
    =-\frac{F_\beta}{F_c}<0,
    \quad
    \frac{\partial \hat c}{\partial T}
    =-\frac{F_T}{F_c}>0.
    \]
    Hence on the interior region \(c^\star=\hat c\) increases strictly in
    \(\lambda\) and \(T\), and decreases strictly in \(\beta\).  
\end{proof}

\subsection{Proof of Proposition~\ref{prop:constant_eq}}
\begin{proof}
    Consider the ordinary differential equation defined for $x \in (c, 1)$:
    \[
        g'(x) = \frac{\lambda g(x) (x - p(g(x)))}{\lambda (1 - x)(x - p(g(x)))+ p'(g(x))},
    \]
    with initial condition $g(c) =0$. 
    For the constant price schedule $p(t) = c$, the derivative $p'(t) = 0$ for all $t \in [0,T]$. 
    Substituting into the ODE derived from Theorem~\ref{thm:VBT_necc}, the equation simplifies to
    \[
        g'(x) = \frac{\lambda g(x)(x - c)}{\lambda (1 - x)(x - c)} = \frac{g(x)}{1 - x}, \quad \text{for } x \in (c, 1).
    \]
    Separating variables and integrating both sides, we obtain
    \[
        \frac{\dd g}{g(x)} = \frac{\dx}{1 - x} \quad \Rightarrow \quad \ln |g(x)| = -\ln|1 - x| + \ln C,
    \]
    for some integration constant \( C > 0 \). 
    Exponentiating both sides gives
    \[
        g(x) = \frac{C}{1 - x}.
    \]
    Applying the initial condition \( g(c) = 0 \), we find that the integration constant \( C = 0 \).
    Therefore, the unique solution is given by 
    \[
        g(x) = 0, \quad \text{for all } x \in (c, 1).
    \]

    We now define the candidate threshold time function $w : [0,1] \to [0,\infty]$ in accordance with the structural requirements stated in Theorem~\ref{thm:VBT_necc}:
    \[
        w(v) = 
        \begin{cases}
            \infty & \text{if } v <c, \\
            0 & \text{if } v \in [c,1],
        \end{cases}
    \]
    where $g(x) = 0 $ is the unique solution to the ODE.
    Note that the values of $w(v)$ for $v < c$ and $v = c$ are determined directly by the necessary structural conditions in Theorem~\ref{thm:VBT_necc}, and for $v \in (c, 1)$, $w(v)$ is given by the unique ODE solution.
    Therefore, $w$ is well-defined and unique on the entire domain $v \in [0,1]$.
    
    To verify the sufficient condition in Theorem~\ref{thm:VBT_suff}, we must show that the interim utility function
    \[
    \Pi^b(\tau; v^*) := (v^* - p(\tau)) \cdot e^{-\lambda \tau (1 - w^{-1}(\tau))}
    \]
    achieves a unique local maximum at $\tau = w(v^*)$ for a fixed $v^* \in [c,1)$. 
    By \eqref{eq:w_inverse}, we have \( w^{-1}(\tau) = c \) for all \( \tau \in [0,T] \).
    Therefore, the interim utility function simplifies to
    \[
    \Pi^b(\tau; v^*) = (v^* - c) \cdot e^{-\lambda \tau (1 - c)}.
    \]
    Since $v^* > c$, the coefficient $(v^* - c) > 0$, and $\Pi^b(\tau; v^*)$ is strictly proportional to the exponential decay function $e^{-\lambda \tau (1 - c)}$, which is strictly decreasing in $\tau \in (0,T]$.
    Differentiating, we obtain
    \[
    \frac{\dd}{\dd\tau} \Pi^b(\tau; v^*) = -\lambda (v^* - c)(1 - c) \cdot e^{-\lambda \tau (1 - c)} < 0 \quad \text{for all } \tau \in (0,T].
    \]
    This implies that $\Pi^b(\tau; v^*)$ is strictly decreasing in $\tau$, and thus attains its unique maximum at $\tau = 0$, which equals $w(v^*)$. 
    Therefore, the condition of Theorem~\ref{thm:VBT_suff} is satisfied.

    Now, we compute the expected waiting time and expected revenue for the seller with the constant price schedule $p(t) = c$. 
    By Proposition~\ref{prop:VBT_seller}, the expected waiting time is given by
    \[
        \bar{\tau}_s =\int_0^T e^{-\lambda t (1 - w^{-1}(t))} \dt= \int_0^T e^{-\lambda t (1 - c)} \dt = \frac{1 - e^{-\lambda T (1 - c)}}{\lambda (1 - c)}.
    \]
    The expected revenue is given by     
    \begin{align*}
        \bar{r}_s &= p(0) - p(T) \cdot e^{-\lambda T (1 - p(T))} + \int_0^T e^{-\lambda t (1 - w^{-1}(t))} \cdot p'(t) \, \dt \\
        & = c - c \cdot e^{-\lambda T (1 - c)} + \int_0^T e^{-\lambda t (1 - c)} \cdot 0 \, \dt  =c(1 - e^{-\lambda T (1 - c)}).
    \end{align*}
    \qedhere
\end{proof}

\subsection{Proof of Theorem~\ref{thm:thick_market}} 
\begin{proof}
    The seller’s utility \( U^s \) is given by \eqref{eq:seller_utility}.
    We aim to maximize: 
    \[
    	\int_0^T \left( p'(t) - \beta \right) e^{-\lambda t (1 - p(t))} \dt.
    \]
    
    To solve this dynamic optimization problem, we use the Euler–Lagrange equation. 
    First, define the Lagrangian as:
    \[
    	L(t, p(t), p'(t)) = \left( p'(t) - \beta \right) e^{-\lambda t (1 - p(t))}.
    \]
    Now, we apply the Euler–Lagrange equation of the form:
    \begin{align} \label{Euler_Lagrange}
	    \frac{\partial L}{\partial p} - \frac{\dd}{\dt} \left( \frac{\partial L}{\partial p'} \right) = 0.
    \end{align}
    First, we compute the following partial derivatives:
    \begin{align*}
        \frac{\partial L}{\partial p} &= \lambda t (p'(t) - \beta) e^{-\lambda t (1 - p(t))}, \quad \frac{\partial L}{\partial p'} = e^{-\lambda t (1 - p(t))}.
    \end{align*}
    Next, we compute the total derivative with respect to \( t \) of \( \frac{\partial L}{\partial p'} \):
    \[
    	\frac{\dd}{\dt} \left( \frac{\partial L}{\partial p'} \right) = \lambda  e^{-\lambda  t (1 - p(t))} \left( 1 - p(t) - t p'(t) \right).
    \]
    Substituting these expressions into \eqref{Euler_Lagrange} gives
    \begin{align*}
    0 
    &= \lambda t (p'(t) - \beta) e^{-\lambda t (1 - p(t))} - \lambda e^{-\lambda t (1 - p(t))} (1 - p(t) - t p'(t)) \\
    &= \lambda e^{-\lambda t (1 - p(t))} \left( t (p'(t) - \beta) + p(t) - 1 + t p'(t) \right).
    \end{align*}
    Consequently, we must have
    \begin{align*}
    	0 &= t (p'(t) - \beta) + (1 - p(t) - t p'(t)) \\ 
    	0 &= 1 - p(t) - \beta t.
    \end{align*}
    Thus, the optimal pricing schedule \( p^\star(t) \) satisfies the linear equation:
    \[
	    p^\star(t) = 1 - \beta t.
    \]
    \qedhere
\end{proof}

\subsection{Proof of Proposition~\ref{prop:linear_eq}}
\begin{proof}
    Consider the ordinary differential equation(ODE) defined for $x \in [p(T), 1)$:
    \[
        g'(x) = \frac{\lambda g(x) (x - p(g(x)))}{\lambda (1 - x)(x - p(g(x))) + p'(g(x))},
    \]
    with initial condition $g(p(T)) = \frac{b - p(T)}{m}$. 
    For the linear price schedule $p(t) = b - mt$, we have $p'(t) = -m$, so the equation simplifies to
    \[
        g'(x) = \frac{\lambda g(x) (x - b + m g(x))}{\lambda (1 - x)(x - b + m g(x)) - m}.
    \]
    To apply the Picard--Lindel\"of theorem, we consider the domain
    \[
        \mathcal{D} := \{ (x, g) \in (p(T), 1) \times [0,T] : p(g(x)) \le x \},
    \]
    which ensures the rationality condition holds throughout. Within $\mathcal{D}$, the right-hand side is continuous in $x$ and locally Lipschitz in $g(x)$, as it is a rational function with polynomial numerator and denominator, and the denominator remains bounded away from zero when $g(x) \in [0, T]$.
    Therefore, by the Picard--Lindel\"of theorem, there exists a unique local solution $g(x)$ near $x = p(T)$ that remains in $\mathcal{D}$. 
    Furthermore, since the solution remains bounded and the right-hand side remains regular throughout, it can be uniquely extended to the entire interval $(p(T), b)$.
    
    We now define the candidate threshold time function $w : [0,1] \to [0,\infty]$ in accordance with the structural requirements stated in Theorem~\ref{thm:VBT_necc}:
    \[
        w(v) = 
        \begin{cases}
            \infty & \text{if } v < p(T), \\
            \frac{b - v}{m} & \text{if } v = p(T), \\
            g(v) & \text{if } v \in (p(T), b),
        \end{cases}
    \]
    where $g$ is the unique solution to the ODE established in the preceding analysis. 
    Note that the values of $w(v)$ for $v < p(T)$ and $v = p(T)$ are determined directly by the necessary structural conditions in Theorem~\ref{thm:VBT_necc}, and for $v \in (p(T), 1)$, $w(v)$ is given by the unique ODE solution.
    Therefore, $w$ is well-defined and unique on the entire domain $v \in [0,1]$.

    To verify the sufficient condition in Theorem~\ref{thm:VBT_suff}, we must show that the interim utility function
    \[
    \Pi^b(\tau; v) := (v - p(\tau)) \cdot e^{-\lambda \tau (1 - w^{-1}(\tau))}
    \]
    achieves a unique local maximum at $\tau = w(v^*)$ for a fixed $v^* \in [p(T),b)$. 
    Suppose that there exists another point $\hat{\tau} \in [0, T]$ with $\hat{\tau} \ne w(v^*)$ such that $\hat{\tau}$ is also a local maximizer of $\Pi^b(\tau; v^*)$. Then, we must have
    \[
    \left. \frac{\dd}{\dd\tau} \Pi^b(\tau; v^*) \right|_{\tau = \hat{\tau}} = 0.
    \]
    However, the first-order condition that characterizes the buyer's optimal purchase time leads to the differential equation
    \[
    \frac{\dw}{\dv} = \frac{\lambda T w(v)(v - p(w(v)))}{\lambda T (1 - v)(v - p(w(v))) + p'(w(v))},
    \]
    with initial condition \( w(p(T)) = \frac{b - p(T)}{m} \), whose solution is unique on \((p(T), b)\) by the Picard--Lindelöf theorem.
    Thus, for a given $v^*$, the corresponding $\tau = w(v^*)$ is uniquely determined. 
    If $\hat{\tau} \ne w(v^*)$ were also a maximizer satisfying the same first-order condition, this would contradict the uniqueness of the ODE solution.
    Therefore, the assumption leads to a contradiction, and we conclude that $\tau = w(v^*)$ is the unique local maximizer of the interim utility function.
    
\end{proof}

\end{document}